\pdfoutput=1
\documentclass{IEEEtran}

\usepackage{tcolorbox}
\usepackage[ruled,linesnumbered]{algorithm2e}
\usepackage{bm}
\usepackage{stackrel}
\usepackage{subfigure}
\usepackage{epsf}
\usepackage{amsmath,amssymb}
\usepackage{graphicx}
\usepackage{bmpsize}
\usepackage{color}
\usepackage{cite}
\usepackage{multirow,tabularx}
\usepackage{ifthen}
\usepackage{epstopdf}
\usepackage{mathtools}
\input{epstopdf.sty}
\usepackage{times}

\usepackage{hyperref}

\input{epsf.sty}
\makeatletter
\def\BState{\State\hskip-\ALG@thistlm}
\makeatother

\newcommand{\hide}[1]{\ifthenelse{\boolean{false}}{#1}{}}

\newtheorem{theorem}{{\bf Theorem}}

\newtheorem{lemma}{{\bf Lemma}}

\newtheorem{remark}{\textit{Remark}}

\newenvironment{definition}[1][Definition]{\begin{trivlist}
\item[\hskip \labelsep {\bfseries #1}]}{\end{trivlist}}

\newcommand{\qed}{\nobreak \ifvmode \relax \else
      \ifdim\lastskip<1.5em \hskip-\lastskip
      \hskip1.5em plus0em minus0.5em \fi \nobreak
      \vrule height0.75em width0.5em depth0.25em\fi}

\DeclareMathOperator*{\argmax}{arg\,max}

\newcommand{\beq}{\begin{equation}}
\newcommand{\eeq}{\end{equation}}
\newcommand{\barr}{\begin{array}}
\newcommand{\earr}{\end{array}}

\newcommand{\benum}{\begin{enumerate}}
\newcommand{\eenum}{\end{enumerate}}

\newcommand{\bit}{\begin{itemize}}
\newcommand{\eit}{\end{itemize}}

\newcommand{\bc}{\begin{center}}
\newcommand{\ec}{\end{center}}

\newcommand{\bdes}{\begin{description}}
\newcommand{\edes}{\end{description}}

\newcommand{\bfig}{\begin{figure}}
\newcommand{\efig}{\end{figure}}

\newcommand{\bemq}{\begin{quote} \begin{em}}
\newcommand{\eemq}{\end{em} \end{quote}}

\newcommand{\bmp}{\begin{minipage}}
\newcommand{\emp}{\end{minipage}}

\newcommand{\secref}[1]{Section~\ref{#1}}

\newcommand{\bsp}{\begin{slide*}}
\newcommand{\esp}{\end{slide*}}
\newcommand{\bsl}{\begin{slide}}
\newcommand{\esl}{\end{slide}}

\newcommand{\blem}{\begin{lemma}}
\newcommand{\elem}{\end{lemma}}
\newcommand{\bthm}{\begin{theorem}}
\newcommand{\ethm}{\end{theorem}}

\graphicspath{{./figures/}}

\IEEEoverridecommandlockouts

\begin{document}

%\title{Minimizing Age-of-Information in Multi-Hop Wireless Networks}
%\title{Scheduling Policies for Age Minimization in Multi-Hop Communication Networks}
%\title{Scheduling Policies for Minimizing Age of Information in Multi-Hop Communication Networks}
%\title{Guaranteeing Information Freshness in Multi-Hop Communication Networks}
%\title{Information Freshness in Multi-Hop Communication Networks}
\title{Information Freshness in Multi-Hop Wireless Networks}
%\date{12 April 2016}
\author{Vishrant Tripathi, Rajat Talak, and Eytan Modiano
\thanks{The authors are with the Laboratory for Information and Decision Systems (LIDS) at the Massachusetts Institute of Technology (MIT), Cambridge, MA. {\tt \{vishrant, talak, modiano\}@mit.edu} \\
}
%\thanks{This work was supported by NSF Grants AST-1547331, CNS-1713725, and CNS-1701964.}
}

% \IEEEaftertitletext{\vspace{-0.6\baselineskip}}

\maketitle

\newcommand{\edge}[2]{\ensuremath{(#1, #2)}}

\newcommand{\Tx}[3]{\ensuremath{U_{#1 #2}^{#3}}}
\newcommand{\CS}[2]{\ensuremath{S_{#1 #2}}}
\newcommand{\CSprob}[2]{\ensuremath{\gamma_{#1 #2}}}

\newcommand{\Age}[2]{\ensuremath{A^{#2}_{#1}}}

\newcommand{\NonLin}[2]{\ensuremath{g^{#1}_{#2}}}
\newcommand{\AgeN}[2]{\ensuremath{B^{#2}_{#1}}}
\newcommand{\Aave}{\ensuremath{A_{\text{ave}}}}
\newcommand{\Bave}{\ensuremath{B_{\text{ave}}}}
\newcommand{\weight}[2]{\ensuremath{w^{#2}_{#1}}}

\newcommand{\red}[1]{\textcolor[rgb]{1.00,0.00,0.00}{#1}}

\newcommand{\event}[2]{\ensuremath{\mathcal{E}_{#1}^{#2}}}
\newcommand\independent{\protect\mathpalette{\protect\independenT}{\perp}}
\def\independenT#1#2{\mathrel{\rlap{$#1#2$}\mkern2mu{#1#2}}}

\newcommand{\linkActFreqSum}[2]{\ensuremath{f_{#1 #2}}\xspace}
\newcommand{\linkActFreq}[3]{\ensuremath{f_{#1 #2}^{#3}}\xspace}
\newcommand{\linkActFreqVec}[1]{\ensuremath{\mathbf{f}^{#1}}\xspace}

\newcommand{\linkFreqSet}{\ensuremath{\mathcal{F}}\xspace}

\newcommand{\EX}[1]{\mathbb{E}\left[ #1 \right]} % expectation operator
\newcommand{\pr}[1]{\mathbb{P}\left[ #1 \right]}
\newcommand{\rt}[1]{\red{RT:~{#1}}}

\newcommand{\setX}{\ensuremath{\mathbb{X}}\xspace}

\newcommand{\edgel}[1]{\ensuremath{e_{#1}}}

\newcommand{\Path}[1]{\ensuremath{p^{#1}}}

\newcommand{\ADweights}[3]{\ensuremath{\Delta_{#1 #2}^{#3}}}

\newcommand{\pos}[1]{\ensuremath{\left[ #1\right]^{+}}}

\begin{abstract}
We consider the problem of minimizing age of information in multihop wireless networks and propose three classes of policies to solve the problem - stationary randomized, age difference, and age debt. For the unicast setting with fixed routes between each source-destination pair, we first develop a procedure to find age optimal \textit{Stationary Randomized} policies. These policies are easy to implement and allow us to derive closed-form expression for average AoI. Next, for the same unicast setting, we develop a class of heuristic policies, called \textit{Age Difference}, based on the idea that if neighboring nodes try to reduce their age differential then all nodes will have fresher updates. This approach is useful in practice since it relies only on the local age differential between nodes to make scheduling decisions. Finally, we propose the class of policies called \textit{Age Debt}, which can handle 1) non-linear AoI cost functions; 2) unicast, multicast and broadcast flows; and 3) no fixed routes specified per flow beforehand. Here, we convert AoI optimization problems into equivalent network stability problems and use Lyapunov drift to find scheduling and routing schemes that stabilize the network. We also provide numerical results comparing our proposed classes of policies with the best known scheduling and routing schemes available in the literature for a wide variety of network settings. 

%\rt{Abstract of Allerton17 paper.}
%Timely exchange of information over multi-hop wireless networks is gaining increasing relevance with growing interests in applications such as internet of things (IoT) and autonomous vehicular networks. Age-of-information (AoI) is a recently proposed performance metric that measures information freshness at the destination node. AoI at a destination node is the time since last update was received.
%
%We study AoI for multi-hop networks with general interference constraints with $\mathcal{R}$ source-destination pairs, and derive simple stationary policies in which links are activated according to a stationary probability distribution. We first consider a line network with a single source-destination pair, and characterize AoI as a convex function of link activation rates. We then use this result to obtain the optimal policy, in the class of stationary policies, for multi-hop network, with several source-destination pairs. We prove an important \emph{separation principle}, which says that the optimal scheduling policy for the multi-hop problem can be obtained by solving an equivalent problem in which all source-destination pairs are single-hop away.
\end{abstract}

\section{Introduction}
\label{sec:introduction}
Emerging applications such as networked control systems, real-time surveillance and monitoring, augmented and virtual reality, cloud gaming, and caching at the wireless edge rely crucially on the continuous delivery of fresh updates over communication networks. Further, exchanging fresh information updates over \textit{multi-hop} wireless networks is gaining increasing relevance with the advent of ad-hoc networked wireless systems such as internet of things (IoT), vehicular networks, and networks of unmanned aerial vehicles. 

%In unmanned aerial vehicular networks, for example, exchanging position, velocity, and other control information in a timely fashion can help in collision avoidance and efficient path planning~\cite{talak16_cdc_speed_limits}. In IoT, and other cyber physical systems, timely feedback of sensor data is vital to the overall system performance.,FANETs2013

These systems differ from the traditional communication systems in two ways. In traditional communication systems, data or packet arrival is assumed to be an exogenous process that cannot be controlled. However, in a lot of real-time applications, the generation of update packets, such as sensor data, can be controlled. It has been shown~\cite{2012Infocom_KaulYates} that generating update packets at the right rate can improve freshness, striking a balance between too high a rate of generation that results in network congestion and too low a rate that results in updates being sent too infrequently.

Secondly, traditional communication systems use packet centric performance measures such as throughput or delay to characterize performance. These performance measures do not fully capture the information freshness paradigm. For example, delay of a stale packet, that got caught in the network due to network clogging, doesn't need to be accounted for as long as the intended ground station gets fresh information regularly via other, promptly received, update packets.

\begin{figure}
  \centering
  \includegraphics[width=0.98\linewidth]{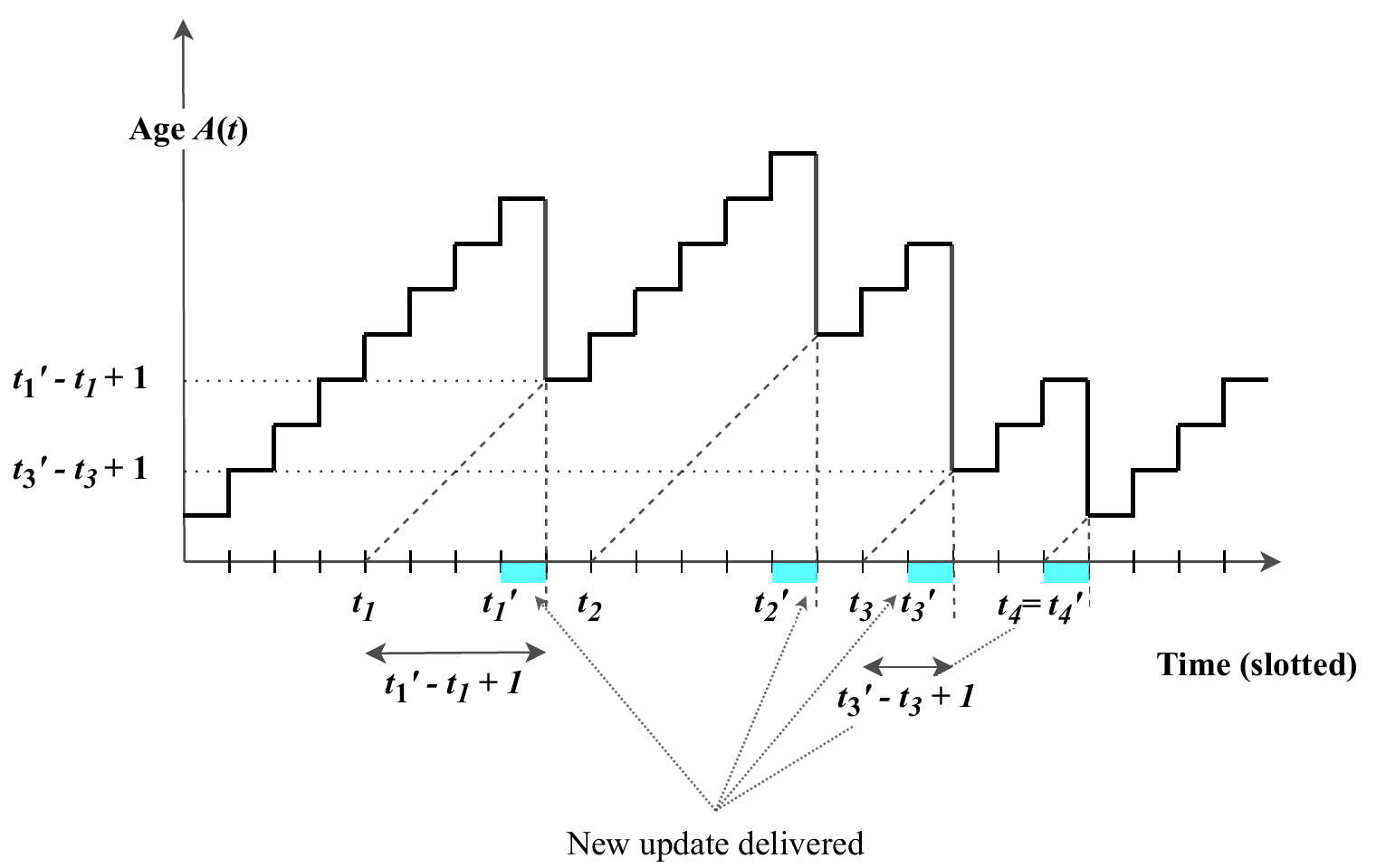}
  \caption{Age of Information (AoI) as a function of time. Here, $t_{i}$ is the time of generation of the $i$th packet at the source, and $t_{i}^{'}$ is the time of its reception at the destination node.}
  \label{fig:age_evolve}
\end{figure}

A new performance measure, called Age of information (AoI), was proposed in~\cite{2011SeCON_Kaul, 2012Infocom_KaulYates} to measure information freshness at the destination node. AoI at the destination node at time $t$, is the time elapsed since the last received update packet was generated. Figure~\ref{fig:age_evolve}, plots AoI evolving in time. Whenever the destination node receives a fresh update packet, the AoI drops to the time elapsed since the received packet's generation time, while it grows linearly otherwise. 

Over the past decade, there has been a rapidly growing body of work analyzing AoI in different queuing models \cite{2012Infocom_KaulYates, 2015ISIT_LongBoEM, 2012CISS_KaulYates, 2016X_Najm, 2014ISIT_CostaEp, 2013ISIT_KamKomEp, 2014ISIT_KamKomEp, 2016X_LongBo, Inoue18_FCFS_LCFS_AoIDist} and as a scheduling metric in single-hop wireless networks \cite{2016allerton_IgorAge, 2017ISIT_YuPin, talak18_ISIT, Igor18_infocom, Igor18_AoI_IndexPolicies, talak18_StGenIC_Mobihoc, talak18_WiOpt}. We review these works briefly in Section \ref{sec:rel_work}. For detailed surveys on AoI literature, we point the reader to \cite{kosta2017age_book} and \cite{sun2019age_book}.

%AoI was first studied in~\cite{2011SeCON_Kaul} for a vehicular network using simulation. Vehicles periodically generated update packets to be transmitted to other nodes in the network. An optimal rate of packet generation was observed. To better understand this phenomena,~\cite{2012Infocom_KaulYates} modeled the network between the source and destination as a single first-in-first-out (FIFO) queue, and proved that there is indeed an optimal rate at which AoI is minimized.
 
Minimizing AoI over multi-hop networks with general interference constraints, however, has received limited attention. In~\cite{2016Ep_WiOpt}, a switch type network was considered under physical interference constraints, and the problem of scheduling finitely many update packets was shown to be NP-hard for this network. AoI in multi-hop networks of queues was studied in~\cite{BedewyISIT17_LIFO_opt}, where LIFO queue service was shown to reduce age. AoI minimization in multihop wireless networks with all-to-all broadcast flows was considered in \cite{farazi_2018_multihop, RickBrown19_InfocomW_AoImultihop}. Scaling of AoI in multihop multicast networks was studied in \cite{Ulukus_2019_multicast}.

Finding low complexity near optimal scheduling and routing schemes for AoI minimization which handle general network topologies, interference constraints, cost functions, different types of flows and link reliabilities has remained an open problem.

\subsection{Contributions}
In this work, we develop a unifying framework for making routing and scheduling decisions that minimize AoI cost in general multihop networks. In Section~\ref{sec:system-model}, we describe the system model for multihop networks with general interference constraints; unicast, multicast and broadcast flows; general non-linear cost functions of AoI; and unreliable links. We consider the problem of minimizing long-term AoI cost over such networks.

In Section~\ref{sec:srp}, we consider the simple class of stationary randomized policies, where scheduling and routing decisions are taken in an i.i.d. manner from a fixed probability distribution. We restrict analysis of this class of policies to weighted-sum AoI minimization over multihop networks with \textit{only} unicast flows and \textit{known, fixed} paths between each source-destination pair, i.e. we only need to make scheduling decisions and not routing decisions. First, we derive a closed form expression for the average AoI for each source-destination pair under any specified randomized policy. We then show that finding the best stationary randomized policy for AoI minimization over multihop networks can be converted into an equivalent single-hop AoI minimization problem. We discuss examples of how to solve this optimization problem and provide performance bounds which suggest that even the best randomized policies can be far from optimal in large networks.

In Section~\ref{sec:age-diff}, we develop a heuristic policy called the \textit{Age Difference} policy, based on the idea that if the age differential between nodes is small, all nodes can get fresh updates and have low AoI. We also restrict the discussion of this policy to weighted-sum AoI minimization over multihop networks with \textit{only} unicast flows and \textit{known, fixed} paths between each source-destination pair. We show that the age difference policy is a myopic policy that greedily optimizes for a specific AoI cost in every time-slot. We further discuss simple examples that illustrate how the age difference policy outperforms all stationary randomized policies.

In Section~\ref{sec:age-debt}, we consider the multihop problem in full generality - 1) with non-linear AoI cost functions; 2) unicast, multicast and broadcast flows and 3) considering both scheduling and routing decisions for optimization. We provide a recipe to transform AoI optimization problems into network stability problems. Instead of trying to solve for the best scheduling and routing policies directly, we assume that we have access to a set of target values which represent the average age cost for every flow in the network. These target values could be application specific freshness requirements provided by a network administrator, or they could be the solution to an optimization program that optimizes some utility function of the average age costs.

In Section~\ref{sec:age-debt-sub}, we introduce the notion of \textit{Age Debt} and set up a virtual queuing network that is stable if and only if there exists a feasible network control policy that can achieve the specified target costs. In Section~\ref{sec:drift}, we use Lyapunov drift based methods to stabilize this system of virtual queues and achieve the desired target age costs. In Section~\ref{sec:c_alpha}, we further discuss how to choose the right age cost targets, when there is no access to either an optimization oracle or a system administrator specifying requirements for each flow. Finally, in Section~\ref{sec:simulations}, we provide detailed simulation results that compare our proposed AoI optimization methods with prior works. We find that Age Debt and its variants perform as well as or better than the best known scheduling and routing schemes in a wide variety of network settings.

\subsection{Related Work}
\label{sec:rel_work}
Age for FIFO M/M/1, M/D/1, and D/M/1 queues was analyzed in~\cite{2012Infocom_KaulYates}, multiclass FIFO M/G/1 and G/G/1 queues were  studied in~\cite{2015ISIT_LongBoEM}, while last-in-first-out (LIFO) queues under various arrival and service time distributions were studied in~\cite{2012CISS_KaulYates, 2016X_Najm, 2014ISIT_CostaEp}. AoI for M/M/2 and M/M/$\infty$ queues was analyzed in~\cite{2013ISIT_KamKomEp, 2014ISIT_KamKomEp}, which primarily studied the impact of out-of-order delivery of packets on age. Effects of packet error or packet drop on age for the M/M/1 queue, with FIFO service, was studied in~\cite{2016X_LongBo}. Closed-form expressions for the stationary distribution of AoI in single-server queues were derived in \cite{Inoue18_FCFS_LCFS_AoIDist}.

More recently, a number of works have also looked at AoI as a metric for designing wireless scheduling policies and solving the problem of minimizing AoI in single-hop wireless networks. In \cite{2016allerton_IgorAge, 2017ISIT_YuPin, talak18_ISIT, Igor18_infocom, Igor18_AoI_IndexPolicies, talak18_StGenIC_Mobihoc, talak18_WiOpt}, the authors consider the problem of minimizing AoI in multiple access type networks with nodes and a single base station, where only a few links can be activated at any given time. These works typically prove constant factor optimality of three classes of policies - randomized, max-weight and Whittle index based; under both reliable and unreliable channels. Slotted ALOHA-like random access for AoI minimization has also been studied in multiple recent works~\cite{2017ISIT_KaulYates_AoI_ALOHA, 2021_Elif_ALOHA, 2021_Igor_ALOHA}. Further, minimizing general non-linear cost functions of AoI in single-hop wireless networks has been considered in \cite{jhunjhunwala_2018_functions,tripathi_2019_whittle}.

\subsection{Prior Versions}
Preliminary versions of this work appeared in Allerton 2017 \cite{talak17_allerton} and in the INFOCOM AoI Workshop 2021 \cite{tripathi_2021_agedebt}. In \cite{talak17_allerton}, we introduced stationary randomized policies for multihop networks while in \cite{tripathi_2021_agedebt}, we introduced the Age Debt policy. This paper combines these two lines of inquiry into a general framework for AoI optimization over multihop wireless networks. In addition, we also propose and analyze a third policy for AoI minimization called \textit{Age Difference} and provide more substantial numerical results.

\section{System Model}
\label{sec:system-model}

Consider a network with $N$ nodes connected by a fixed undirected graph $G(V,E)$. An edge $\edge{i}{j}$ means that nodes $i$ and $j$ can send packets to one another directly. 
We assume that at most one update can be sent over an edge in any given time-slot and takes exactly one time-slot to get delivered. We normalize the time-slot duration to unity.
%

%
%We denote by $\mathcal{S}$ the set that contains every set of possible one-hop packet forwarding actions on graph $G$ that can be scheduled simultaneously without interference. A feasible network control policy chooses routing and scheduling actions only from this set $\mathcal{S}$. %An example of a typical interference constraints is only allowing edges with no destination nodes in common to transmit at the same time.

\textbf{Flows.} The network consists of $K$ ($\leq N$) source nodes that generate information updates. All the sources are active, i.e. they generate fresh updates on demand. A source node $k$ has to send these updates to a set of $D_k \subset N$ destination nodes in the network. We assume that a set of nodes $C_k \subset N$ is commissioned for each flow $k$ to forward its update packets to destination nodes. For example, a network administrator could restrict the paths over which certain flows are allowed.
A flow is characterized by the triplet of source node, its commissioned nodes, and the destination nodes, namely $(k, C_k, D_k)$. For simplicity, we use $k$ to denote both the source node $k$ and the flow corresponding to source node $k$. Note that a node could be both a destination node and also a commissioned node forwarding packets for flow $k$, i.e. $C_k \cap D_k$ is not necessarily empty. 

%Thus, each flow can be identified by its unique source node. 
Flows can be of three types depending on the number of destination nodes: (1) unicast: the flow has a single destination node. (2) multicast: the flow has multiple destination nodes, which are a strict subset of the remaining nodes. (3) broadcast: every node other than the source itself is a destination node. The commissioned nodes can be either a small subset of nodes in the network needed to reach all the destination nodes, or the entire network.
We assume there to be no queuing at any node and that each node maintains a single packet buffer for the freshest packet of each flow. \footnote{ Discarding older packets, or equivalently, preemptive LCFS (last come first serve) is known to be the optimal queuing discipline for AoI minimization \cite{2016_ISIT_TIT_YinSun_Thput_Delay_LCFS}.}

\textbf{Interference and Link States.} We consider unreliable links as well as general interference constraints, i.e., transmission on all the links cannot happen simultaneously. We enumerate the set of all possible interference free choices of links and corresponding flow transmissions in the set $\mathcal{A}$. Thus, a member of set $\mathcal{A}$ contains a subset of links and corresponding flows which can be sent on these links in a single time-slot without interference. A valid network control policy must choose an action that is a member of the set $\mathcal{A}$ in every time-slot. Note that this description of $\mathcal{A}$ is very general and allows for interference constraints that depend on flow assignments. For example, consider a setting where a node is allowed to broadcast updates of a single flow to all of its neighbors in a single timestep but not send updates regarding \textit{different} flows to each neighbor simultaneously.

%We call $m \subset E$ a \emph{feasible set} if transmission can occur on all edges in $m$ successfully, without interference. We let $\mathcal{A}$ denote the collection of all feasible sets.
%
For link $\edge{i}{j} \in E$, we use \Tx{i}{j}{k}(t) and \CS{i}{j}(t) (both $\in \{0, 1\}$) to denote the transmission decision and link state of the link \edge{i}{j} at time $t$. \Tx{i}{j}{k}(t) is $1$ if a transmission of a flow-$k$ update is scheduled on the link, at time $t$, and is $0$ otherwise.
Whereas, \CS{i}{j}(t) is $1$ if a scheduled transmission on the link, at time $t$, will succeed; provided there is no interference. We assume $\{ \CS{i}{j}(t) \}_{t, (i,j)}$ to be independent and identically distributed processes across time $t$ and links $(i,j)$, with $\CSprob{i}{j} = \pr{\CS{i}{j}(t)=1}$.

\textbf{Age Evolution.} For a flow $k$, each commissioned and destination node keeps track of the age of the freshest packet it has received. For a node $j \in C_k\cup D_k$, we denote its age for the $k$th flow by \Age{j}{k}(t) and it evolves as:
\begin{equation}
\label{eq:AoI_evolution_mh}
\Age{j}{k}(t+1) = \left\{ \begin{array}{ll}
                    \min(\Age{j}{k}(t),\Age{i}{k}(t)) + 1 &\text{if}~\Tx{i}{j}{k}(t)\CS{i}{j}(t) = 1 \\
                    \Age{j}{k}(t)+1, &\text{if}~\Tx{i}{j}{k}(t)\CS{i}{j}(t) = 0
                  \end{array} \right.,
\end{equation}
for all $i \in \{ k\} \cup C_k$, $j \in C_k\cup D_k$, and link $\edge{i}{j} \in E$. Note that for any flow, the source node and the commissioned nodes transmit the update packets, while other commissioned nodes and the destination nodes receive them.

%For every destination node $j \in D_k$, for flow $k$, we maintain an age of information process $\Age{j}{k}(t)$ which tracks how old the information is at node $j$ about node $k$.
%\begin{equation}
%\label{eq:AoI_evolution_mh}
%\Age{j}{k}(t+1) =
%\begin{cases}
%\begin{aligned}
%\min(\Age{j}{k}(t),t - t_g) + 1, \text{if update generated}\\
%\text{at time }t_g\text{ is delivered at time }t.
%\end{aligned}\\
%\Age{j}{k}(t)+1, \text{if no new delivery at time }t.
%\end{cases}
%\end{equation}

\textbf{Information Freshness.} We consider two metrics of information freshness. The first is the average weighted sum AoI at the destination nodes:
\begin{equation}\label{eq:aveAoI}
\Aave = \lim_{T \rightarrow \infty} \EX{\frac{1}{T}\sum_{t=1}^{T} \sum_{k=1}^{K}\sum_{j \in D_k} \weight{j}{k} \Age{j}{k}(t)},
\end{equation}
where \weight{j}{k} denote constant weights, which determines the relative importance of a destination $j$ and flow $k$, with respect to others.
For the second metric, we consider general possibly non-linear functions of age. We associate a monotone increasing age cost function for each source-destination pair $(k, j)$, where $j \in D_k$,  denoted by $\NonLin{k}{j}(\cdot)$. We define the non-linear, effective age process to be:
\begin{equation}
\AgeN{j}{k}(t) \triangleq \NonLin{k}{j}(\Age{j}{k}(t)),
\end{equation}
for all $t \geq 1$.
%Using these age cost functions, we maintain the effective age processes $\AgeN{j}{k}(t) \triangleq \NonLin{k}{j}(\Age{j}{k}(t))$. 
The non-linear age metric is defined as:
\begin{equation}\label{eq:aveFAoI}
\Bave = \lim_{T \rightarrow \infty} \EX{\frac{1}{T}\sum_{t=1}^{T} \sum_{k=1}^{K}\sum_{j \in D_k} \AgeN{j}{k}(t) },
\end{equation}
which is a generalized version of \Aave~in~\eqref{eq:aveAoI}.

Our goal is to minimize either \Aave~or \Bave~for a general, multi-hop network, by determining a policy that controls the link transmissions. A control policy needs to specify not only which links should be scheduled in each time-slot but also which flows should be transmitted along each link. We assume a centralized controller. 

%%% NOT SURE IF WE NEED THIS
%We enumerate the set of all possible interference free choices of links and corresponding flows in the set $\mathcal{S}$. Thus, a member of set $\mathcal{S}$ contains a subset of links and corresponding flows which can be sent on these links in a single time-slot without interference. A valid network control policy must choose an action that is a member of the set $\mathcal{S}$ in every time-slot.%Note that this description of $\mathcal{S}$ also allows for interference constraints that depend on flow assignments such as when a node is allowed to broadcast a single packet to all of its neighbors but not send different flow packets to each of them.
%This problem can be written as:
%%The typical goal of AoI-based scheduling and routing design for multihop networks is to minimize the time average of the expected age costs summed across flows:
%\begin{equation}
%\label{eq:age_cost_opt_mh}
%\pi^{*} = \underset{\pi}{\operatorname{argmin}}~~~\Bave,
%\end{equation}
%where $\pi(t) \in \mathcal{S}, \forall t,\pi$.

In the next three sections, we propose three different policies, in the order of their complexity and performance, to minimize the information freshness metrics.

\section{Stationary Randomized Policy}
\label{sec:srp}

In this section, we look at the simplest case of our general multi-hop model. We consider a class of policies called stationary randomized policies, in which each feasible action $a \in \mathcal{A}$ is activated with a given probability, independently across time. The analysis of these policies is limited to the setting where each flow is \textit{unicast} and consists of a \textit{single known path} to forward updates from the source to each destination. We derive the exact expression of the average age metric in this case and show how to optimize it over the interference constraints.

\subsection{Stationary Randomized Policies}
\label{sec:stat_policy}
We first define the space of stationary randomized policies over a multi-hop network. Let \event{\edge{i}{j}}{k}(t) be the event that a transmission over link \edge{i}{j} is attempted, to send updates of flow $k$, at time-slot $t$.
\begin{definition}\label{def:st_policy}
$\pi$ is a \emph{stationary randomized policy} if the events $\event{\edge{i}{j}}{k}(t)$ and $\event{\edge{i}{j}}{k}(t')$ are independent and stationary for any $t \neq t'$, for all links \edge{i}{j} and flows $k$. Further
\begin{equation*}
%\event{\edge{i}{j}}{k}(t) \independent \event{\edge{i}{j}}{k}(t')~\text{and}~
\pr{\event{\edge{i}{j}}{k}(t)} = \pr{\event{\edge{i}{j}}{k}(t')} = \linkActFreq{i}{j}{k},
\end{equation*}
for all time-slots $t$ and $t'$. Here $\linkActFreq{i}{j}{k}$ is the frequency of occurrence of the event \event{\edge{i}{j}}{k}. %and also independent across flow $k$, namely,
%\begin{equation*}
%\event{\edge{i}{j}}{k}(t) \independent \event{\edge{i}{j}}{k'}(t),
%\end{equation*}
%for all flows $k$ and $k'$.
\end{definition}

Note that all stationary randomized policies are associated with probabilities \linkActFreq{i}{j}{k}. We refer to \linkActFreq{i}{j}{k} as the \emph{link activation frequency} of link \edge{i}{j} for flow $k$, and  use \linkActFreqVec{k} to denote the tuple $\{ \linkActFreq{i}{j}{k} \}_{\edge{i}{j} \in E}$.

A way to generate the space of all stationary randomized policies possible with a centralized scheduler is the following: transmit on all links the corresponding flow choices for an action $m \in \mathcal{A}$, with probability $x_m$, independently across time-slots $t$. The probabilities $x_m$ can be varied to produce different stationary randomized policies, but are naturally constrained by the fact that they must sum to unity, namely, $\sum_{m \in \mathcal{A}} x_m = 1$. We use $\setX$ to denote the space of $\mathbf{x} = (x_m)_{m \in \mathcal{A}}$ such that $x_m$s sum to unity.

This induces the link activation frequency $\linkActFreq{i}{j}{k}$  given by
\begin{equation}
\label{eq:x_to_f}
\linkActFreq{i}{j}{k} = \sum_{ \{m\in \mathcal{A}:~(\Tx{i}{j}{k} = 1)\in m \} } x_m,
\end{equation}
%and the $k$th flow link activation frequency $\linkActFreq{i}{j}{k}$ given by
%\begin{equation}
%\label{eq:f_to_fk}
%\linkActFreq{i}{j}{k} = %q^{k}_{ij}\linkActFreqSum{i}{j},
%\end{equation}
for all links \edge{i}{j} and flows $k$. The above equation simply states that the link activation frequency $\linkActFreq{i}{j}{k}$ is the sum of the activation probabilities of all actions $m \in \mathcal{A}$ in which a flow $k$ packet is transmitted over link $(i,j)$, i.e. $\Tx{i}{j}{k} = 1$. We use $\linkActFreqVec{} = M\mathbf{x}$ to denote~\eqref{eq:x_to_f} and \linkFreqSet to denote the space of all feasible link activation frequencies $\{ \linkActFreqVec{k} \}_{k} = \linkActFreqVec{}$.
We will see that this space plays a critical role in determining the stationary randomized policy that minimizes average age.

\textbf{Single-hop Age Problem.} Consider the special case when we are interested in minimizing \Aave~in a single hop network, with general interference constraints and multiple nodes sending updates to a base station (BS). Since each edge simply connects one node to the BS, it can only forward packets from that node. Thus, the notation for the link activation frequencies can be simplified from $\linkActFreq{i}{j}{k}$ to $\linkActFreq{e}{}{}$ where $e$ is the only edge connecting node $e$ to the BS and also the only edge transmitting flow $e$ packets.
It is easy to see that for such a network, the weighted average age minimization problem can be written as:%~\eqref{eq:shap}.
\begin{align}
\label{eq:shap}
\begin{aligned}
& \underset{\mathbf{x} \in \setX}{\text{Minimize}}
& & \sum_{e \in E} \frac{w_e}{\CSprob{e}{}\linkActFreq{e}{}{} }, \\
%& \text{subject to} & &  \sum_{k = 1}^{K} \linkActFreq{\edgel{j}}{}{k} \leq \linkActFreqSum{\edgel{j}}{}~\forall e \in E \\
&\text{subject to} && \mathbf{f} = M \mathbf{x}.
\end{aligned}
\end{align}
We, therefore, call it the single-hop age problem. We will see that all the average age minimization problems in this section - even though we are dealing with multi-hop flows - can be reformulated to look like~\eqref{eq:shap}. In the next section, we derive a simple expression for the average age for a line network under a stationary randomized policy.

\subsection{Line Network}
\label{sec:line_network}
We now analyze the average age for a line network with a single source and a single destination, assuming general interference constraints. 
Consider the line network $G = (V, E)$, where $V = \{1, \ldots N\}$ and
$E = \{ (1,2), (2, 3), \ldots (N-1, N)\}$ denote the $N$ nodes and $N-1$ links, respectively. 
For convenience, we use \edgel{j} to denote the link \edge{j-1}{j} and \Tx{\edgel{j}}{}{} and \CS{\edgel{j}}{} to denote the transmission and channel state on \edge{j-1}{j}. %for all $l = 1, 2, \ldots N-1$.
The network contains a single flow with source node $s = 1$ and this flow has a destination node $D_1 = \{ N\}$. The commissioned nodes include all other nodes, to forward updates from the source to the destination, i.e. $C_1 = \{2, 3, \ldots N-1\}$. The source $s$ generates fresh update packets that are transmitted over the line network to reach the destination node $N$.

For this simple line network, we show that the age evolution in~\eqref{eq:AoI_evolution_mh} can be simplified.
\begin{lemma}\label{lem:age_evol}
The age evolution of~\eqref{eq:AoI_evolution_mh} can be written as
\begin{equation}\nonumber
A_{j}(t+1) = \left\{ \begin{array}{ll}
                       \!\!\!A_{j-1}(t) + 1 &\text{if}~\Tx{e_j}{}{}(t)\CS{e_j}{}(t) = 1  \\
                       \!\!\!A_{j}(t)   + 1 &\text{otherwise}
                     \end{array}\right.,
\end{equation}
for all $j \in C_1 \cup D_1 = \{2, 3, \ldots N\}$.
\end{lemma}
\begin{IEEEproof}
If $\Tx{\edgel{j}}{}{}(t)\CS{\edgel{j}}{}(t) = 0$, then no successful transmission occurs over link \edgel{j}, and therefore, it follows from~\eqref{eq:AoI_evolution_mh} that $\Age{j}{}(t+1) = \Age{j}{}(t) + 1$.
If $\Tx{\edgel{j}}{}{}(t)\CS{\edgel{j}}{}(t) = 1$, then a successful transmission does occurs over link \edgel{j} at time $t$, and two possibilities arise: either node $j-1$ has a fresh update, received since the last transmission over the link \edgel{j}, or it hasn't.
If the node $j-1$ has a fresh update, then this fresh update packet is transmitted to node $j$ at time $t$, and we get $\Age{j}{}(t+1) = \Age{j-1}{}(t) + 1$. %Since $A_{l}(t) \geq A_{l-1}(t)$ for all $t$, this is same as $A_{l}(t+1) = \min\{A_{l-1}(t), A_{l}(t)\} + 1$.
If, on the other hand, node $j-1$ hasn't received a fresh update since the last transmission over link \edgel{j}, then we have $\Age{j}{}(t) = \Age{j-1}{}(t)$. This is because both nodes have the same update packet that was exchanged during the last transmission over link \edgel{j}. The age evolution, in the absence of a fresh update,  therefore becomes $\Age{j}{}(t+1) = \Age{j}{}(t) + 1 = \Age{j-1}{}(t) + 1$.
\end{IEEEproof}

The age evolution equation in Lemma~\ref{lem:age_evol} is true irrespective of the scheduling policy. We now focus on stationary randomized policies described in Section~\ref{sec:stat_policy}. In Section~\ref{sec:stat_policy}, we saw that every stationary policy $\pi$ is associated with a link activation frequencies $\linkActFreq{i}{j}{k}$. For the stationary randomized policies, we now characterize the average age at the destination node $N$ as a function of the link activation frequencies.
\begin{theorem}
\label{thm:age_lineNet}
If $\linkActFreq{e}{}{} > 0$ is the link activation frequency for link $e \in E$ under a stationary policy $\pi$ then the average age at node $N$ is given by
\begin{equation}
\label{eq:no1}
\Aave = \lim_{T \rightarrow \infty} \EX{ \frac{1}{T}\sum_{t=1}^{T} \Age{N}{}(t) } %= \sum_{j=2}^{N}\frac{1}{\CSprob{\edgel{j}}{}\linkActFreq{\edgel{j}}{}{}},
= \sum_{e \in E} \frac{1}{\CSprob{e}{}\linkActFreq{e}{}{}},
\end{equation}
where \CSprob{e}{} denotes the channel state probability for link $e \in E$.
%where \CSprob{\edgel{j}}{} denotes the channel state probability for link \edgel{j}, for $j \in \{2, 3, \ldots N \}$.
\end{theorem}
\begin{IEEEproof}
See Appendix~\ref{pf:thm:age_lineNet}.
\end{IEEEproof}

Firstly, note that the average age equals 
\begin{equation}
\Aave = \frac{1}{\CSprob{\edgel{1}}{}\linkActFreq{\edgel{1}}{}{}},
\end{equation}
when there is just one link $\edgel{1}$ in the network, i.e., $N=2$ nodes. Theorem~\ref{thm:age_lineNet} shows that the average age for a line graph splits as a sum of expressions of link activation frequencies and link state probabilities, each of which can be thought of as an average age expression for the corresponding link. This, although surprising, happens due to the i.i.d. nature of attempting transmissions in any stationary randomized policy.

%\rt{Comment about how the age splits across links.}

\subsubsection{Average Age Minimization}
Using the result in Theorem~\ref{thm:age_lineNet} we can now formulate the average age minimization problem over a line network, under general interference constraints. This is given by:
%\begin{tcolorbox}\textbf{Line-Network Age Problem}
\begin{align}\label{eq:age_opt_line}
\begin{aligned}
& \underset{\mathbf{x} \in \setX}{\text{Minimize}} & & \sum_{e \in E} \frac{1}{\CSprob{e}{}\linkActFreq{e}{}{} }, \\
& \text{subject to} && \mathbf{f} = M\mathbf{x}.
\end{aligned}
\end{align}
%where  \CSprob{e}{} is the channel state probability for link $e$ and $\mathbf{x}$ denotes the probability distribution over the feasible sets $m \in \mathcal{A}$. 
This is same as the single-hop age problem in~\eqref{eq:shap} with equal link weights $w_e$. 
Note that the complexity of solving~\eqref{eq:age_opt_line} is determined by the interference constraints that are encoded in the matrix $M$ in~\eqref{eq:age_opt_line}.
%\end{tcolorbox}
%
% Notice that we have ignored the flow weight $w^{1}$ as the network contains only one flow, i.e., $\mathcal{R} = \{1\}$.
%
%The set $\mathcal{F}$ is the space of feasible link activation frequencies which depend on the interference constraints, and the type of policy. As we saw in Section~\ref{sec:stat_policy}, that the space $\mathcal{F}$ can be significantly different for centralized and distributed policies, respectively. We discuss the solvability of~\eqref{eq:age_opt_line} in the next section, where we consider age minimization for a general network.

%\rt{Comment how this is similar to the single-hop age problem.}

\subsection{General Network}
\label{sec:gen_network}
We now consider the average age minimization problem with $K$ flows. We assume a single destination node for every flow. Each source node $k$ is assigned a set of nodes $C_k$ in the network that forms a single connected path from the source to the destination node $d_k$. We use \Path{k} to denote the set of all links on the source-destination path induced by nodes $\{k\} \cup C_k \cup \{ d_k\}$. The goal is to determine the optimal stationary randomized scheduling policy that minimizes the weighted average age in~\eqref{eq:aveAoI}. For simplicity, we denote the weight for each source-destination pair $(k,d_k)$ by just $w^{k}$, since we are only consider unicast flows.

%In this section, we consider the problem of age minimization for a general network, where we have a network $G = (V, E)$ with a collection of feasible activation sets $\mathcal{A}$, and $\mathcal{R} = \{1, 2, \ldots R\}$ flows. Given that every flow is assigned a path we obtain the optimal stationary scheduling policy.

%\subsubsection{Multi-Hop Age Problem}
%\label{sec:multi-hop}
%
%We now consider the general multi-hop age minimization problem. Where the paths $p(r)$ may be more than a single-hop away.
%

Let $\pi$ be a stationary randomized policy with link activation frequencies $\{ \linkActFreqVec{k} \}_{k}$.  Applying Theorem~\ref{thm:age_lineNet}, we can write the average age for flow $k$ to be:
\begin{equation}
\Aave^{k} = \lim_{T \rightarrow \infty} \EX{\frac{1}{T}\sum_{t=1}^{T} \Age{d_k}{k}(t)} = \sum_{e \in \Path{k} } \frac{1}{ \CSprob{e}{} \linkActFreq{e}{}{k} }.
\end{equation}
This is because the set of nodes $\{ k\} \cup C_k \cup \{ d_k\}$ form a path from the source $k$ to the destination node $d_k$. The weighted average age (in~\eqref{eq:aveAoI}) can be written as:
\begin{align}
\Aave   &= \lim_{T \rightarrow \infty} \EX{ \frac{1}{T}\sum_{t=1}^{T} \sum_{k =1}^{K} w^{k}\Age{d_k}{k}(t)}, \\
        &= \sum_{k \in [K]} \sum_{e \in \Path{k}} \frac{w^{k}}{\CSprob{e}{}\linkActFreq{e}{}{k} }.
\end{align}
The average age minimization problem can be written as:
%
%\begin{tcolorbox}\textbf{General-Network Age Problem:}
\begin{align}
\label{eq:age_opt}
\begin{aligned}
& \underset{\mathbf{x} \in \setX}{\text{Minimize}}
& & \sum_{k \in [K]} \sum_{e \in \Path{k}} \frac{w^{k}}{\CSprob{e}{}\linkActFreq{e}{}{k} }, \\
%& \text{subject to} & &  \sum_{k = 1}^{K} \linkActFreq{\edgel{j}}{}{k} \leq \linkActFreqSum{\edgel{j}}{}~\forall e \in E \\
&\text{subject to} && \mathbf{f} = M \mathbf{x}.%\linkActFreqVec{k} = Q^{k}\mathbf{f}~\forall~k = 1, 2, \ldots K,~\text{and}\\
%&&& \mathbf{f} = M \mathbf{x}.
\end{aligned}
\end{align}
%\end{tcolorbox}
%

%%%%%%%%%%%%%%%%%%%%%%%%%%%%%%%%%%%
%
%\rt{You are here.} Let $f_{e}^{r}$ denote the fraction of times link $e$ activates successfully to transmit update of source-destination pair $r$. If $e \notin p(r)$ then $f_{e}^{r} = 0$. Since $f_e$ is the net link activation frequency, we have
%\begin{equation}
%f_e = \sum_{r \in \mathcal{R}, e \in p(r)} f_{e}^{r},
%\end{equation}
%for all $e \in E$.
%%
%%

We now observe that~\eqref{eq:age_opt} can be viewed as an equivalent single-hop age problem of the form~\eqref{eq:shap}. 

\begin{lemma}
\label{lem:equi}
The multihop average age minimization problem over stationary randomized policies given by \eqref{eq:age_opt}) can be viewed as an equivalent single-hop age problem of the form \eqref{eq:shap}.
%\begin{equation}
%w_{e} = \left( \sum_{k=1}^{K} \sqrt{w^{k}} \mathbb{I}_{e \in \Path{k}} \right)^{2},
%\end{equation}
%for all links $e \in \bigcup_{k=1}^{K} \Path{k}$ and $w_e = 0$, otherwise. If $\mathbf{f}$ are the optimal link activation frequencies to this single-hop age problem, then $\linkActFreqVec{k}$s are given by
%\begin{equation} \label{eq:flow_assign}
%\linkActFreq{e}{}{k} = \left[ \frac{\sqrt{w^{k}}}{\sum_{k'=1}^{K} \sqrt{w^{k'}}\mathbb{I}_{\{ e \in \Path{k'} \}} }\right] \linkActFreq{e}{}{},
%\end{equation}
%for all $e \in \Path{k}$ and $k = 1, 2, \ldots K$.
\end{lemma}
\begin{IEEEproof}
See Appendix~\ref{pf:thm:equi}.
\end{IEEEproof}
%\color{blue} Comment on how to solve this problem for single-hop and line networks.\color{black}
%
%
%
%Theorem~\ref{thm:equi} proves an important \emph{separation principle} in  designing stationary randomized policies for age minimization over general network. It states that the optimal stationary randomized policy can be obtained by converting all the flows to single-hop, with suitable link weights.
%%
%Another implication of Theorem~\ref{thm:equi} is that given the link activation frequency $f^{\ast}_{e}$, the per-flow frequency $f^{r\ast}_{e}$ can be determined locally.

In this section, we developed a procedure to find age optimal stationary randomized policies in settings with known single commissioned paths between sources and destinations. These policies are easy to implement and analyze. They further allow us to derive closed form expressions of average age and provide performance guarantees. However, as we will see in the coming sections, these policies can only be used in limited settings and their performance is significantly far from optimal in practice.

\section{Age Difference Policy}
\label{sec:age-diff}

We  now discuss a heuristic policy that yields a much lower average age, in practice, than even the optimal stationary randomized policy. We are still restriced to the setting with only unicast flows and known commissioned paths for each source-destination pair.

The basic idea is as follows: \emph{in the propagation of information updates, it is important to keep the age differential between two neighboring nodes to as small a value as possible.} %If this gap is large, then the receiver node can gain fresh information. %We, therefore, propose the following policy.

\textbf{Example.} To illustrate this, consider a line network with $N$ nodes and $N-1$ links, with the single source and destination node placed at the two ends of the network. Assume all other nodes are commissioned for sending updates from the source to the destination node. Let the probability of successful transmission $\CSprob{e}{}$ be $1$ and the interference constraint be such that a transmission can occur only on one of the $N-1$ links, at any given time.

For this network, we can deduce from Section~\ref{sec:srp} that the average age, given by $\Aave = \sum_{j =2}^{N} \frac{1}{f_{\edgel{j}}}$, is minimized at $f_{\edgel{j}} = 1/(N-1)$ for all $j$ and equals $\Aave^{\ast} = (N-1)^2 = \mathcal{O}(N^2)$.

Now, consider a scheduling policy that works to minimize the age differential between any two nodes that share a link. This is done by scheduling the link \edgel{j} that has the maximum age differential. We schedule link \edgel{j^{\ast}}, at time $t$, such that
\begin{equation}\label{eq:ADexample}
j^{\ast} = \argmax_{j} \pos{\Age{j}{}(t) - \Age{j-1}{}(t)}.
\end{equation}
It can be deduced that under this scheduling policy over the line network the age at the destination node records the periodic pattern: $\{N-1, N, N+2, \ldots 2(N-1), N-1, N, N+2, \ldots 2(N-1), N-1, N, \ldots \}$. Thus, the average age at the destination node can be computed to be $\Aave = (N-1) + N/2 = \mathcal{O}(N)$. Note that this is a significant improvement over the average age optimal stationary randomized policy.

\textbf{Age-Difference Policy.} We now articulate this age-difference heuristic for the general multi-flow, multi-hop network.
%
%\begin{definition}
Let $\ADweights{i}{j}{k}(t)$ be the age difference weight for link $\edge{i}{j}$ and flow $k$ given by
\begin{equation}\label{eq:ADweight}
\ADweights{i}{j}{k}(t) \triangleq \weight{j}{k}\CS{i}{j}(t)\pos{ \Age{j}{k}(t) - \Age{i}{k}(t) },
\end{equation}
for all $j \in C_k \cup D_k$, $i \in \{k\}\cup C_k$, and flow $k \in [K]$. Note that $\weight{j}{k}$ denote the flow $k$ weights for all the commissioned and the destination nodes. The $\weight{j}{k}$s, for the destination nodes $j \in D_k$, were defined in the average age cost function in~\eqref{eq:aveAoI}.  For all commissioned nodes, these weights are set to $1$, i.e. 
\begin{equation}\nonumber
\weight{j}{k} = \left\{ \begin{array}{ll}
\!\!\!\weight{j}{k} &\text{if}~k \in D_k  \\
\!\!\! 1 &\text{otherwise}
\end{array}\right..
\end{equation}
Note that this choice is heuristic, and other sets of weights might lead to slightly different scheduling behavior. 

From \eqref{eq:ADweight}, we note that the age-difference policy can utilize information about the channel states $\CS{i}{j}(t)$ when known, unlike the stationary randomized policy which utilizes edges irrespective of whether the channel is currently on or off. When channel states $\CS{i}{j}(t)$ for links are unknown, the age-difference weight is given by replacing $\CS{i}{j}(t)$ with the average reliability $\gamma_{ij}$ for link $(i,j)$. 
%\begin{equation}\label{eq:ADweight2}
%\ADweights{i}{j}{k}(t) \triangleq {\weight{j}{k}}\gamma_{ij}\pos{ \Age{j}{k}(t) - \Age{i}{k}(t) },
%\end{equation}
%Further, let
%\begin{equation}\label{eq:ADweightSum}
%\ADweights{i}{j}{}(t) \triangleq \sum_{k = 1}^{K} \ADweights{i}{j}{k}(t).
%\end{equation}

The \emph{age-difference} policy %does the following operation in each time-slot $t$:
%\begin{enumerate}
  %\item 
  schedules a feasible set $m(t)$ that maximizes the age-difference weight, namely:
        \begin{equation}\label{eq:AD_feasibleSet}
        m(t) = \argmax_{m \in \mathcal{A}} \sum_{(e,k) \in m} \ADweights{e}{}{k}(t).
        \end{equation}
%  \item For each link $e \in m(t)$, it selects a flow $k_{e}(t)$ such that 
%        \begin{equation}\label{eq:AD_flowSelect}
%        k_{e}(t) = \argmax_{k \in [K]} \ADweights{e}{}{k}(t),
%        \end{equation}
%        and attempts transmission of the information update on link \edge{i}{j} corresponding to flow $k_{e}(t)$.
%\end{enumerate}
%\end{definition}

\textbf{Result.} We now show that the age-difference policy is in fact a one-step greedy policy for the average age minimization problem in~\eqref{eq:aveAoI}, albeit with a modified age cost per time-slot.
\begin{lemma}
\label{lem:age-diff}
The age-difference policy is a myopic greedy policy in minimizing the average age
\begin{equation}\label{eq:aveAoI_plusCk}
\Aave = \lim_{T \rightarrow \infty} \EX{\frac{1}{T}\sum_{t=1}^{T} \sum_{k=1}^{K}\sum_{j \in C_k\cup D_k} \weight{j}{k} \Age{j}{k}(t)}.
\end{equation}
\end{lemma}
\begin{IEEEproof}
See Appendix~\ref{pf:lem:age-diff}.
\end{IEEEproof}

The average age in~\eqref{eq:aveAoI_plusCk}, differs from the average age defined in~\eqref{eq:aveAoI}, in that it takes into account the age of also the commissioned nodes, for every flow $k$.

Greedy/myopic policies with similar structures have been shown to be constant factor optimal in the single-hop AoI minimization setting in \cite{Igor18_AoI_IndexPolicies, talak18_Mobihoc}. 
%\rt{Make a comments on the use and importance of myopic policies in solving MDPs. Comment that the age-difference policy is known to be constant factor optimal in single-hop average age minimization scenarios~\cite{talak, igor}.}

%\rt{Comment that the age-difference policy makes use of the channel state information, while the stationary randomized policy does not. In the age-difference policy you can replace \CS{i}{j} with \CSprob{i}{j}, in the case of unknown channel states, and the conclusion about it being a myopic policy will hold true.}

%\rt{Can you show that age-difference policies are not optimal in a simple example? It would be a good bridging point to the next section.}
The age-difference policy overcomes a key limitation of the stationary randomized policy, i.e. better performance in practice. However, it is still restricted to settings with a) only unicast flows, and b) source-destination paths that are fixed and known beforehand, and c) weighted sum AoI cost.

In the following section, we develop a general policy design framework that can address settings without any of these limitations.

\section{Age Debt Policy}
\label{sec:age-debt}

In this section, we develop the age-debt framework for AoI minimization, based on ideas from Lyapunov optimization. To do so,
we first introduce the notions of age-achievability and age debt virtual queues. We then show how stabilizing this network of virtual queues leads to minimization of AoI. Finally, we use quadratic Lyapunov drift to propose a heuristic scheme to achieve this stabilization in general multi-hop networks.

Note that in this section, we consider settings with a) general increasing cost functions of AoI, b) no knowledge of fixed routing paths, i.e. the scheduler also needs to make routing decisions and c) unicast, multicast and broadcast flows in the same network. The general AoI optimization problem can be formulated as:
\begin{equation}
\label{eq:age_cost_opt_mh}
\pi^{*} = \underset{\pi}{\operatorname{argmin}} \bigg( \lim_{T \rightarrow \infty}  \mathbb{E} \bigg[\frac{1}{T} \sum_{t = 1}^{T} \sum_{k=1}^{K} \sum_{j \in D_k} B^{k}_{j}(t) \bigg] \bigg),
\end{equation}
where $B^{k}_{j}(t)$ are the effective age processes and $\pi(t) \in \mathcal{A}, \forall t$. This setting is more general than the ones considered in \secref{sec:srp} and \secref{sec:age-diff}. 

\subsection{Age Debt}
\label{sec:age-debt-sub}

We start by assuming that we have been given a target value of time average age cost for each source-destination pair; denoted by $\alpha^{k}_{j}$ for the source-destination pair $(k,j)$. We aggregate the target values associated with each source-destination pair in the vector $\bm{\alpha}$. For any such target vector $\bm{\alpha}$, we define the notion of age-achievability below.
%\begin{framed}
	\begin{definition}
		A vector $\bm{\alpha}$ is \textbf{age-achievable} if there exists a feasible network control policy $\pi$ such that		
		\begin{equation}
		\lim_{T \rightarrow \infty}  \frac{1}{T} \sum_{t = 1}^{T} B^{k}_{j}(t)  \leq \alpha^{k}_{j}, \forall j \in D_k, \forall k \text{ w.p. 1.}
		\end{equation}			
	\end{definition}
%\end{framed}
In other words, a vector $\bm{\alpha}$ is age-achievable if the time-average of the effective age process for \textit{every} source-destination pair $(k,j)$ is upper bounded by the target value $\alpha_{kj}$, under some feasible network control policy. %The expectation of the age process is taken over the randomness in link reliabilities and the scheduling policy.

Note that the combination of general cost functions and achievability targets allows us to capture very general freshness requirements which might be useful in practical system specifications. For example, if an application requires that the empirical distribution of the age process $A^{k}_{j}(t)$ should satisfy $\mathbb{P}(A^{k}_{j}(t) \geq M) \leq \epsilon$, then we can capture this by setting the cost function $g^{k}_{j}(h) = \bm{1}_{ \{h \geq M\}}$ and the corresponding target to be $\alpha^{k}_{j} = \epsilon$. %More complicated constraints on the empirical probability distribution or moments of the age processes can also be specified using cost functions and targets similarly.

We now define a set of virtual queues called age-debt queues for every source-destination pair $(k,j)$. These queues measure how much the effective age process exceeds its target value $\alpha_{kj}$, summed over time. Our definition of debt is inspired by the notion of throughput debt as introduced in \cite{hou_2009_throughput_debt}.
%\begin{framed}
	\begin{definition}
		Given a target vector $\bm{\alpha}$, the \textbf{age debt queue} for source-destination pair $(k,j)$ at time $t$, given by $Q^{k}_{j}(t)$, evolves as 		
		\begin{equation}
		\begin{aligned}
		Q^{k}_{j}(t+1) = \bigg[Q^{k}_{j}(t) + B^{k}_{j}(t+1) - \alpha^{k}_{j}\bigg]^{+}, \forall j \in D_k, \\ \text{ and } \forall k \in \{1,...,K\}.
		\end{aligned}
		\end{equation}
		To complete the definition, each age debt queue starts at zero, i.e. $Q^{k}_{j}(0) = 0, \forall j,k$.		
	\end{definition}
%\end{framed}
We now introduce a notion of stability for these age debt queues. This is similar to how rate stability is typically defined in queueing networks \cite{neely_2010_stability_definitions}.
%\begin{framed}
	\begin{definition}
		We say that the network of age debt queues is \textbf{stable} under a policy $\pi$ and a given target vector $\bm{\alpha}$ if the following condition holds:
		\begin{equation}
		\lim_{T \rightarrow \infty} \mathbb{E} \bigg[\sum_{k=1}^{K} \sum_{j \in D_k} \frac{Q^{k}_{j}(T)}{T} \bigg] = 0,
		\end{equation}
		where the expectation is taken over the randomness in the channel processes and the scheduling policy $\pi$.
	\end{definition}
%\end{framed}
We also establish an equivalence relationship between age-achievability of a vector $\bm{\alpha}$ and the stability of the corresponding network of age debt queues.
%\begin{framed}
\begin{lemma}
	\label{lem:debt_equi}
	A target vector $\bm{\alpha}$ is age-achievable \textit{if and only if} there exists a network control policy $\pi$, that stabilizes the network of source-destination age debt queues.
\end{lemma}
\begin{IEEEproof}
	See Appendix \ref{pf:lem_equi}.
\end{IEEEproof}

%\end{framed}
Next, we define a debt-stable scheduling policy. Such a policy takes a target vector $\bm{\alpha}$ as an input and stabilizes the network of corresponding age debt queues.
%\begin{framed}
	\begin{definition}
		A \textbf{debt-stable} scheduling policy $\pi$ stabilizes the set of age-debt queues for any given target vector $\bm{\alpha}$ that is age-achievable.
	\end{definition}
%\end{framed}

The notions introduced until now effectively allow us to convert the minimum age cost problem described by \eqref{eq:age_cost_opt_mh} into a network stability problem. Suppose $\pi^{*}$ is a solution to the optimization problem \eqref{eq:age_cost_opt_mh}. Further, suppose that the time average of the effective age process for pair $(k,j)$ under $\pi^{*}$ is given by %$\alpha^{*}_{kj}$, i.e.
\begin{equation}
\lim_{T \rightarrow \infty}  \mathbb{E} \bigg[\frac{1}{T} \sum_{t = 1}^{T} B^{k^*}_{j}(t) \bigg] = \alpha^{k^*}_{j}, \forall (k,j).
\end{equation}
%Then, a debt-stable policy will also be age cost optimal when given the vector $\bm{\alpha^{*}}$ as an input.
Clearly, if we have oracle access to an optimal age cost vector $\bm{\alpha^{*}} = \{\alpha^{k^*}_{j}\}_{(k,j)}$ and know how to design a debt-stable policy then we can perform minimum age cost scheduling. If the debt-stable policy is much lower in computational complexity than solving \eqref{eq:age_cost_opt_mh} directly, then we can also solve \eqref{eq:age_cost_opt_mh} at the same lower complexity (assuming oracle access to $\bm{\alpha^{*}}$). We now discuss a heuristic approach to designing debt-stable policies.

\subsection{Lyapunov Drift Approach}
\label{sec:drift}
\subsubsection{Single-Hop Broadcast}
\label{sec:single_hop}
We first consider the special case of single-hop broadcast networks. This setting is easier to analyze since it only requires scheduling and no routing and it also highlights key structural properties of our proposed policy.

Consider a $N$ node star network where each of the nodes $1,...,N-1$ has an edge to node $N$. These nodes wish to send packets to the central node $N$. The edges are numbered $e_1, ..., e_{N-1}$. Due to broadcast interference constraints, only one node can transmit in any given time-slot. Since the destination for every flow is $N$, we can drop the destination in our notation. The age evolution is given by
\begin{equation}
\label{eq:AoI_evolution}
A_{i}(t+1) =
\begin{cases}
A_{i}(t)+1, & \text{if } U_{e_i}(t)S_{e_i}(t) = 0 \\
1, &  \text{if }U_{e_i}(t)S_{e_i}(t) = 1.
\end{cases}
\end{equation}
%Here $\pi(t)$ is the source scheduled in time-slot $t$ and $c_i(t)$ is an indicator variable denoting edge reliability between node $i$ and node $N$ at time $t$.
Given an age-cost function $g_{i}(A_{i}(t))$ and a corresponding target value $\alpha_i$, the debt queue evolution for node $i$ is given by:
\begin{equation}
\label{eq:s_debt}
Q_{i}(t+1) = \bigg[Q_{i}(t) + g_i(A_{i}(t+1)) - \alpha_i\bigg]^{+}. %\triangleq \sum_{k=1}^{t} (B^{\pi}_i(k) - \alpha_i).
\end{equation}

Given a target vector $\bm{\alpha}$, we will use a Lyapunov drift based scheduling scheme to try and achieve debt stability. To do so, we first define a Lyapunov function for our system of virtual queues:
\begin{equation}
L(t) \triangleq \sum_{i=1}^{N-1} Q^{2}_i(t).
\end{equation}
% Using this Lyapunov function, we then define the Lyapunov drift:
% \begin{equation}
%     \Delta(t+1) \triangleq L(t+1) - L(t).
% \end{equation}

Using this Lyapunov function, we then define the \textit{age debt scheduling policy} $\pi^{\text{AD}}$ as:
\begin{equation}
\label{eq:age_debt_policy}
\pi^{\text{AD}}(t) = \underset{a \in \mathcal{A}}{\operatorname{argmin}} \bigg( \mathbb{E} \big[ L(t+1) - L(t)\big] \bigg),
\end{equation}
where the expectation is taken over the randomness in channel reliabilities $\bm{S}(t)$. 

In the following remark, we consider a variant of the age-debt policy that minimizes an upper-bound on the Lyapunov drift instead of the actual Lyapunov drift as in \eqref{eq:age_debt_policy}. This is similar to the upper-bound drift minimization used in policies such as max-weight \cite{Neely_2006_MW_monograph} and allows us to compare the structure of age-debt to preivously proposed policies in literature.
\begin{remark}
	\label{lem:drift_exp}
	Suppose that the links between each source $i$ and the destination $N$ are i.i.d. Bernoulli w.p. $\gamma_i$ in every time-slot. Further, if each age cost function $g_i(\cdot)$ is upper bounded by a large constant $D$, then the policy $\pi(t)$ below minimizes an upper bound on the Lyapunov drift in every time-slot.
	\begin{equation}
	\pi(t) = \underset{i \in 1,...,N-1}{\operatorname{argmax}} \bigg( \gamma_i Q_i(t) \big(g_i(A_i(t)+1) - g_i(1)\big) \bigg).
	\end{equation}
\end{remark}
\begin{IEEEproof}
	See Appendix \ref{pf:drift}.
\end{IEEEproof}
In other words, an approximate drift minimizing policy chooses the source with the largest product of link reliability, current age debt and current age cost. This structure of the age-debt policy can be contrasted with the max-weight policy proposed in \cite{Igor18_AoI_IndexPolicies} which chooses the source with the largest value of $\gamma_i w_i A_i(t) (A_i(t)+2)$ given weights $w_i$. Similarly, the Whittle index policy proposed in \cite{tripathi_2019_whittle}, chooses the source with the largest value of $W_i(A_i(t))$, where $W_i(\cdot)$ is Whittle-index corresponding to the age cost $f_i(\cdot)$.

Note that to compute $\pi^{\text{AD}}(t)$, the scheduler needs to iterate over the set of sources only once. So the per slot computational complexity of this policy grows linearly in $N$. This is similar to the complexity of the Whittle index policy proposed in \cite{Igor18_AoI_IndexPolicies,tripathi_2019_whittle} and the max-weight policies proposed in \cite{Igor18_AoI_IndexPolicies,talak19_ToN_spp}. By contrast, a dynamic programming approach to solve \eqref{eq:age_cost_opt_mh} directly has per slot computational complexity that grows exponentially in $N$. This highlights the key strength of our approach. If the scheduler has some way to set the targets for each source optimally, then the age debt policy is a good low complexity heuristic for age minimization.

\subsubsection{General Networks}
\label{sec:multihop}

The general multihop setting is more challenging. Simply using one-slot Lyapunov drift to try and achieve debt stability does not work directly in the multihop setting. We highlight this with a simple example.

\begin{figure}
	\centering
	\includegraphics[width=0.7\linewidth]{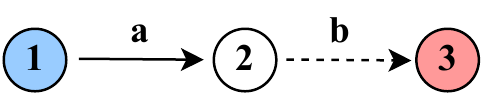}
	\caption{Example of a line network with a unicast flow from node 1 to node 3}
	\label{fig:two_hop}
\end{figure}

Consider the three node network described in Figure~\ref{fig:two_hop} with a single unicast flow from node $1$ to node $3$. The interference constraint enforces that only one of the two edges $a$ and $b$ can be activated in any time-slot. Suppose that we are interested in minimizing the time average of the age process $A^{1}_{3}(t)$. Given a target value $\alpha^{1}_{3}$, we set up the age debt queue as follows:
\begin{equation}
Q^{1}_{3}(t+1) = \bigg[Q^{1}_{3}(t) + A^{1}_{3}(t+1) - \alpha^{1}_{3}\bigg]^{+}.
\end{equation}

We will try to use the one slot Lyapunov drift minimizing policy to stabilize $Q^{1}_{3}(t)$ in this network. To do so, we solve the following optimization in every time-slot:
\begin{equation}
\pi^{\text{AD}}(t) = \underset{x \in \{a,b\} }{\operatorname{argmin}} \bigg( \mathbb{E} \big[ (Q^{1}_{3}(t+1))^{2} - (Q^{1}_{3}(t))^{2}\big] \bigg).
\end{equation}

At $t=1$, activating either edge $a$ or edge $b$ has no effect on the debt $Q^{1}_{3}(2)$ since node $2$ does not have any packet from node $1$. If we break ties in favour of edge $b$, then it is activated but no new packet is delivered to node $3$. At $t=2$, since node $2$ still does not have any new update from node $1$, no action taken can affect the debt $Q^{1}_{3}(3)$. Using the same tie-break rule, we would again schedule edge $b$. This process keeps on repeating and the age debt queue $Q^{1}_{3}$ blows up irrespective of the value of $\alpha^{1}_{3}$, even though the age optimal policy in this setting is to simply alternate between $a$ and $b$ in every time-slot.

The example above illustrates why one-slot Lyapunov drift based techniques fail in stabilizing debt queues in multihop networks. The policy designer using Lyapunov drift is constrained to optimizing \textit{only one time-step into the future}, similar to a greedy policy. So, if every possible scheduling and routing action has no effect on the age debt queues in the immediate next time-slot, the one step drift minimizing procedure does not provide any information on which action should be chosen to stabilize the debt queues.

This suggests that to be able to use one-slot drift minimizing techniques for stability, there should be a virtual queue for every intermediate node that tracks both the current age debt at the destination and the potential reduction in debt at the destination upon forwarding a fresh packet. If we can set up such queues, then large values of debt at intermediate nodes would lead to fresh packets being sent to the next hops via one-slot drift minimizing actions, eventually reaching the destination and stabilizing the age debt queues. %This is akin to the backpressure policy for network stabilization

Let $Q^{k\rightarrow i}_{j}(t)$ denote such a debt queue corresponding to flow $(k,j)$ at an intermediate node $i$. These additional queues at every intermediate node combined with the original debt queues form our virtual network. The Lyapunov function that we use for scheduling and routing is given by:
\begin{equation}
L(t) \triangleq \sum_{k=1}^{K} \sum_{j \in D_k} \bigg((Q^{k}_{j}(t))^{2} + \sum_{i \notin D_k, i \neq k} (Q^{k\rightarrow i}_{j}(t))^2 \bigg)
\end{equation}

The Age Debt scheduling and routing policy is to choose the activation set and corresponding flows that minimizes the expected Lyapunov drift.
\begin{equation}
\pi^{\text{AD}}(t) = \underset{a \in \mathcal{A}}{\operatorname{argmin}} \bigg( \mathbb{E} \big[ L(t+1) - L(t)\big] \bigg),
\end{equation}
where the expectation is taken over the randomness in channel reliabilities $\bm{S}(t)$.

\subsubsection{Intermediate Debt Queues}

We now discuss how to set up the age debt queues $Q^{k\rightarrow i}_{j}(t)$ for intermediate nodes to augment the original network of queues. Note that there are no intermediate nodes for broadcast flows since every node other than the source is a destination. %For unicast or multicast flows, not all nodes in the network are destination nodes. We specify how to set up virtual debt queues for these intermediate forwarding nodes.

Consider a source-destination pair $(k,j)$ for a unicast/multicast flow $k$ and an intermediate node $i$ that is not a destination for the flow originating at $k$. We want to set up the age debt queue $Q^{k\rightarrow i}_{j}(t)$ at $i$ for the pair $(k,j)$. We maintain an age process for flow $k$ at node $i$, even though there is no associated cost or target value for this age process.
\begin{equation}
A^{k}_{i}(t+1) =
\begin{cases}
\begin{aligned}
\min(A^{k}_{i}(t),t - t_g) + 1, \text{if update generated}\\
\text{at time }t_g\text{ is delivered at time }t.
\end{aligned}\\
A^{k}_{i}(t)+1, \text{if no new delivery at time }t.
\end{cases}
\end{equation}
Here $A^{k}_{i}(t)$ measures how old the information at node $i$ is regarding node $k$. We split the debt queue's evolution into two cases.

\textbf{Case 1}: When node $i$ forwards a flow $k$ packet on a set of adjacent links $L$. Let $h^{L}_{ij}$ be the minimum number of hops it takes to reach node $j$ from node $i$, where the first hop can only include edges in the set $L$. Here, $h^{L}_{ij}$ measures the minimum delay with which the packet that was forwarded by $i$ gets delivered at $j$. The age debt queue $Q^{k\rightarrow i}_{j}(t)$, when node $i$ is forwarding a flow $k$ packet along the link set $L$, evolves as:%Using $h^{L}_{ij}$ and the target value at the destination $\alpha_{kj}$
\begin{equation}
\label{eq:c1}
\begin{aligned}
Q^{k\rightarrow i}_{j}(t+1) = \bigg[Q^{k\rightarrow i}_{j}(t) + g^{k}_{j}\big(\min \{A^{k}_{i}(t),A^{k}_{j}(t) \} \\ + h^{L}_{ij}\big) - \alpha^{k}_{j}\bigg]^{+}.
\end{aligned}
\end{equation}
This measures the most optimistic change in age debt possible at the destination using the current packet transmission from node $i$.

\textbf{Case 2:} When node $i$ does not forward a packet from node $k$ along any of its adjacent edges, then the age debt queue evolves as below.
\begin{equation}
\label{eq:c2}
Q^{k\rightarrow i}_{j}(t+1) = \bigg[Q^{k\rightarrow i}_{j}(t) + B^{k}_{j}(t+1) - \alpha^{k}_{j}\bigg]^{+}.
\end{equation}
This means that the intermediate queue simply tracks the change in debt at the destination when it is not forwarding a relevant packet. If the destination is not receiving fresh packets from anywhere in the network then this would increase the intermediate debt queue.

Thus, the debt at an intermediate node $i$ for a source-destination pair $(k,j)$ blows up if (a) either the destination has not received fresh packets for a long time and node $i$ did not forward any packets from $k$ (i.e. \eqref{eq:c2}) or if (b) node $i$ keeps forwarding stale packets from $k$ (i.e. \eqref{eq:c1}). A drift minimizing policy will then try to ensure that either the destination debt queue is small, or node $i$ forwards fresh packets of flow $k$ towards the destination.

%\subsubsection{Breaking Ties}

\subsection{Choosing Target Vectors}
\label{sec:c_alpha}
In the preceding sections, we have developed a general framework of age achievability where given a target average age cost for every source-destination pair, we formulate a corresponding network stability problem and attempt to solve it via one slot Lyapunov drift minimization. In this section, we discuss how to choose the right target vectors, such that they lead to minimum sum age cost.% or possibly optimize some general utility functions of age cost. %We view this problem as an analogue of the network utility maximization framework where a utility function of long term throughput rates needs to be optimized for a network subject to stability of the underlying queuing network.

In the absence of an optimization oracle that provides access to $\bm{\alpha^{*}}$ or a system administrator who specifies average age cost targets based on the underlying application requirements, we develop a simple heuristic to dynamically update $\bm{\alpha}$ in order to optimize utility based on the state of the underlying debt queues.

The following optimization problem needs to be solved to find the best target vector $\bm{\alpha^{*}}$.
\begin{equation}
\label{eq:alpha_opt}
\begin{aligned}
\underset{\bm{\alpha}}{\operatorname{argmin}} & ~\bigg( \sum_{k=1}^{K} \sum_{j \in D_k} \alpha^{k}_{j} \bigg),\\
\text{s.t. } & \bm{\alpha} \text{ is age-achievable}.
\end{aligned}
\end{equation}
Note that this problem has the same optimal value as \eqref{eq:age_cost_opt_mh}.

\subsubsection{Gradient Descent}
We want to use a gradient descent like approach to solve \eqref{eq:alpha_opt} and find $\bm{\alpha^{*}}$. The problem with doing so is that we do not have a simple characterization of the age-achievability region or a low complexity method to test whether a vector is achievable or not. %In fact, we do not even know if the region is convex.

To resolve this, we use Lemma \ref{lem:debt_equi}. If the network of source-destination age debt queues is unstable for a given value of $\bm{\alpha}$, then $\bm{\alpha}$ lies outside the age-achievability region. This immediately suggests the gradient descent like approach described in Algorithm \ref{alg:ADGD}.

\begin{algorithm}
	\DontPrintSemicolon
	%\SetAlgoLined
	%\KwResult{Write here the result}
	\SetKwInOut{Input}{Input}\SetKwInOut{Output}{Output}
	\Input{epoch size $W$, number of epochs $E$, step-size $\eta > 0$, threshold $\epsilon > 0$, initialization $\bm{\alpha}{(1)}$}
	%\Output{Write here the output}
	\BlankLine
	%Set $\bm{\alpha}_0 \leftarrow 0$ \\		
	\While{ $e \in 1,...,E$ }{
		Set up age debt queues using $\bm{\alpha}{(e)}$ and initialize each queue to 0\;
		\While{ $t \in 1,..., W$}{
			Schedule and route using age debt $\pi^{\text{AD}}(t) = \underset{a \in \mathcal{A}}{\operatorname{argmin}} \bigg( \mathbb{E} \big[ L(t+1) - L(t)\big] \bigg)$,
		}		
		\If{$\exists \text{ flow } k\text{ and }j \in D_k$ s.t. $Q^{k}_{j}(W) > \epsilon W$}{
			Increase target values for unstable queues:\\	
			$\alpha^{k}_{j}(e+1)  = \alpha^{k}_{j}(e) + \eta$, $\forall (k,j)$ s.t. $Q^{k}_{j}(W) > \epsilon W$\;
			Other targets remain unchanged:\\
			$\alpha^{k}_{j}{(e+1)}  = \alpha^{k}_{j}{(e)}$, $\forall (k,j)$ s.t. $Q_{kj}(W) \leq \epsilon W$\;
		}		
		\Else{
			%\If{$Q_{kj}(W) \leq \epsilon W, \forall (k,j)$}{
			Update all target values using gradients:	
			$\alpha^{k}_{j}{(e+1)}  = \alpha^{k}_{j}{(e)} - \eta$, $\forall (k,j)$.\;% \nabla g_{kj}(\alpha^{(e)}_{kj}) $, $\forall (k,j)$.\;
		}
	}
	\caption{Age Debt - Gradient Descent}
	\label{alg:ADGD}
\end{algorithm}

The algorithm above runs the age debt policy for epochs of length $W$ time-slots. Within an epoch, the target vector $\bm{\alpha}$ remains fixed. At the end of the epoch, we use the value of the source-destination age debt queues $Q^{k}_{j}(\cdot)$ to update the corresponding targets. If the network has at least one queue with debt larger than a threshold, it suggests that the current vector is not achievable. So, we increase the values of $\bm{\alpha}$ for the source-destination pairs with large values of debt. If the network has all queues with debt below a threshold, the current vector is likely achievable. So, we update the entire target vector using gradient descent. Note that this approach takes a large number of time-slots to converge to a good candidate target vector $\bm{\alpha}$.

\subsubsection{Flow Control}
Another way to dynamically set the target vectors is to take a flow control approach for solving the optimization problem \eqref{eq:alpha_opt}, similar to \cite{Neely_2006_MW_monograph}. Algorithm \ref{alg:ADFC} describes the details.
\begin{algorithm}
	\DontPrintSemicolon
	%\SetAlgoLined
	%\KwResult{Write here the result}
	\SetKwInOut{Input}{Input}\SetKwInOut{Output}{Output}
	\Input{parameter $V>0$, upper bound $\alpha_{\text{max}}$, initialization $\bm{\alpha}{(1)}$}
	%\Output{Write here the output}
	\BlankLine
	%Set $\bm{\alpha}_0 \leftarrow 0$ \\		
	\While{ $t \in 1,...,T$ }{
		Use $\bm{\alpha}{(t)}$ to update debt queue values at time $t$ \;
		Update $\bm{\alpha}$ by solving the optimization below:\\
		\begin{equation*}
		\bm{\alpha}{(t+1)} =
		\begin{aligned}
		\underset{\bm{\alpha}}{\operatorname{argmin}} & \bigg( \sum_{k=1}^{K} \sum_{j \in D_k} V \alpha^{k}_{j} - \alpha^{k}_{j} Q^{k}_{j}(t) \bigg)  ,\\
		\text{s.t.  } & \bm{\alpha} \geq 1, \bm{\alpha} \leq \alpha_{\text{max}}.
		\end{aligned}
		\end{equation*}\;
		Use $\bm{\alpha}{(t+1)}$ to compute the scheduling and routing decision that minimizes drift:
		$\pi(t) = \underset{a \in \mathcal{A}}{\operatorname{argmin}} \bigg( \mathbb{E} \big[ L(t+1) - L(t)\big] \bigg)$\;						
	}		
	\caption{Age Debt - Flow Control}
	\label{alg:ADFC}
\end{algorithm}

The flow control based age debt policy tries to tradeoff between the stability of the queueing network and the optimization of targets using a parameter $V > 0$. In every time-slot, the flow control optimization sets the target $\bm{\alpha}$ for the next time-slot and then the scheduling and routing decisions are computed by minimizing Lyapunov drift.

%In the special case when the utility functions are identity, i.e. $g_{kj}(x) = x, \forall (k,j)$,
The update optimization in step 4 of Algorithm \ref{alg:ADFC} can be simplified to the rule below:
\begin{equation}
\alpha^{k}_{j}{(t+1)} = \begin{cases}
\alpha_{\text{max}}, &\text{ if } Q^{k}_{j}(t) > V\\
1, &\text{ if }Q^{k}_{j}(t) \leq V,
\end{cases}
\forall (k,j).
\end{equation}
Thus, instead of converging to a target vector as in the case with gradient descent,
the flow control approach dynamically switches the value of targets in every time-slot. This means we do not need to wait a long period of time for convergence. When current debts are high, future targets are set to be high pushing the debts lower. Similarly, when the current debts are low, future targets are also set low, pushing the debts higher. The parameter $V$ decides the threshold between high and low values of the debt queues.

\section{Numerical Results}
\label{sec:simulations}
\begin{figure}
	\centering
	\includegraphics[width=0.99\linewidth]{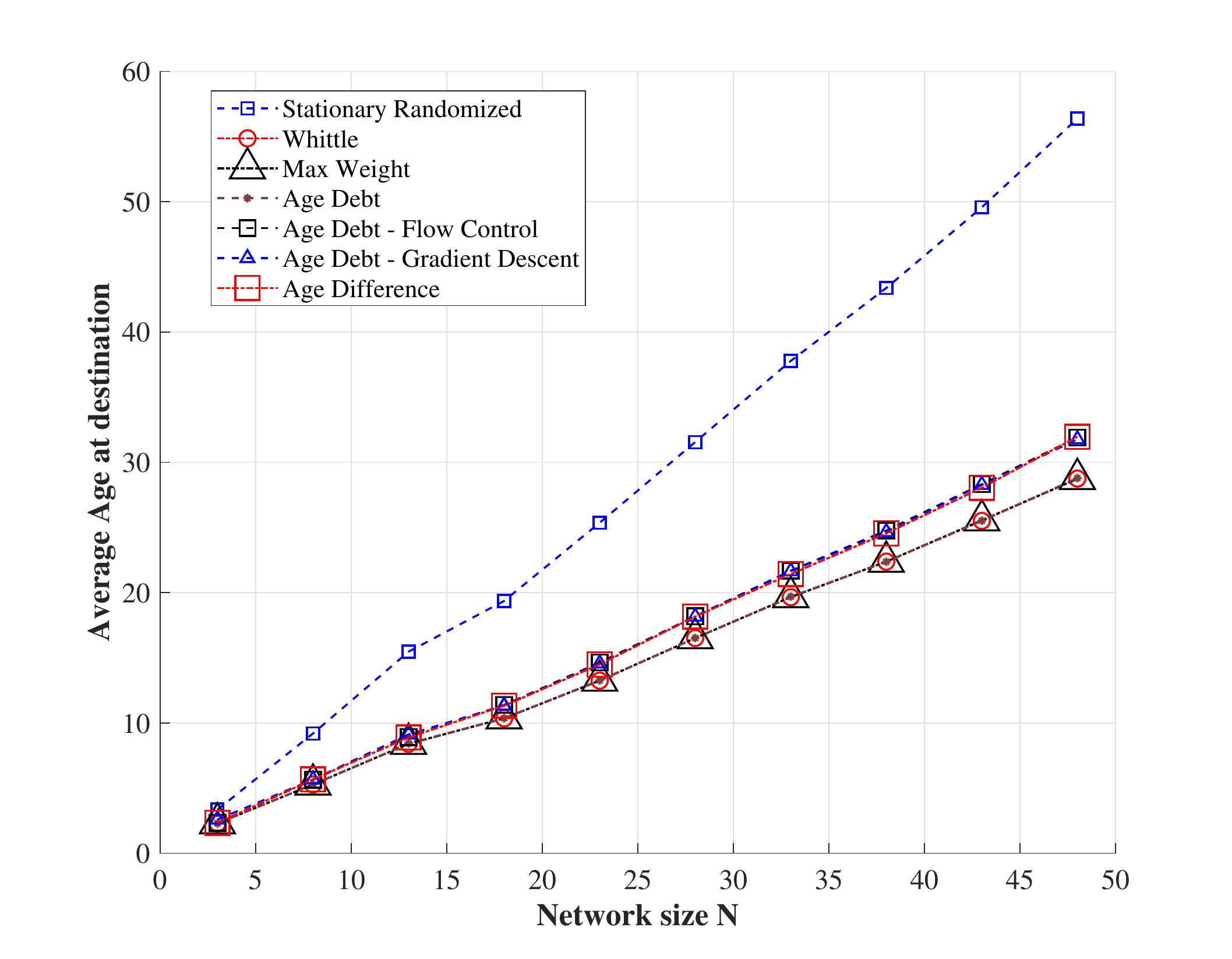}
	\caption{Weighted-sum AoI minimization in broadcast networks with unreliable channels}
	\label{fig:single_lin_uch}
\end{figure}
First, we consider the weighted-sum AoI problem in single-hop broadcast networks with unreliable channels. There are $N$ nodes in the network and the weight of the $i$th node $w_i$ is set to $i/N$. Link connection probabilities are chosen uniformly from the set $[0.6,1]$. Figure \ref{fig:single_lin_uch} plots the performance of the age debt policy, the age difference policy, and the optimal stationary randomized policy along with the max-weight and Whittle index policies proposed in \cite{Igor18_AoI_IndexPolicies} which are known to be close to optimal. 

First, we observe that the optimal stationary randomized policy performs much worse than the other classes of policies. Age difference performs better than the randomized policy but not as well as the Whittle-index or max-weight policies.
We further observe that when the age debt policy is provided the max-weight average cost as the target vector, it replicates near optimal performance. Also, the flow control and gradient descent versions of age debt have a small gap to the max-weight/Whittle policies despite not having access to $\bm{\alpha}$ beforehand and perform as well as the age-difference policy.

\begin{figure}
	\centering
	\includegraphics[width=0.99\linewidth]{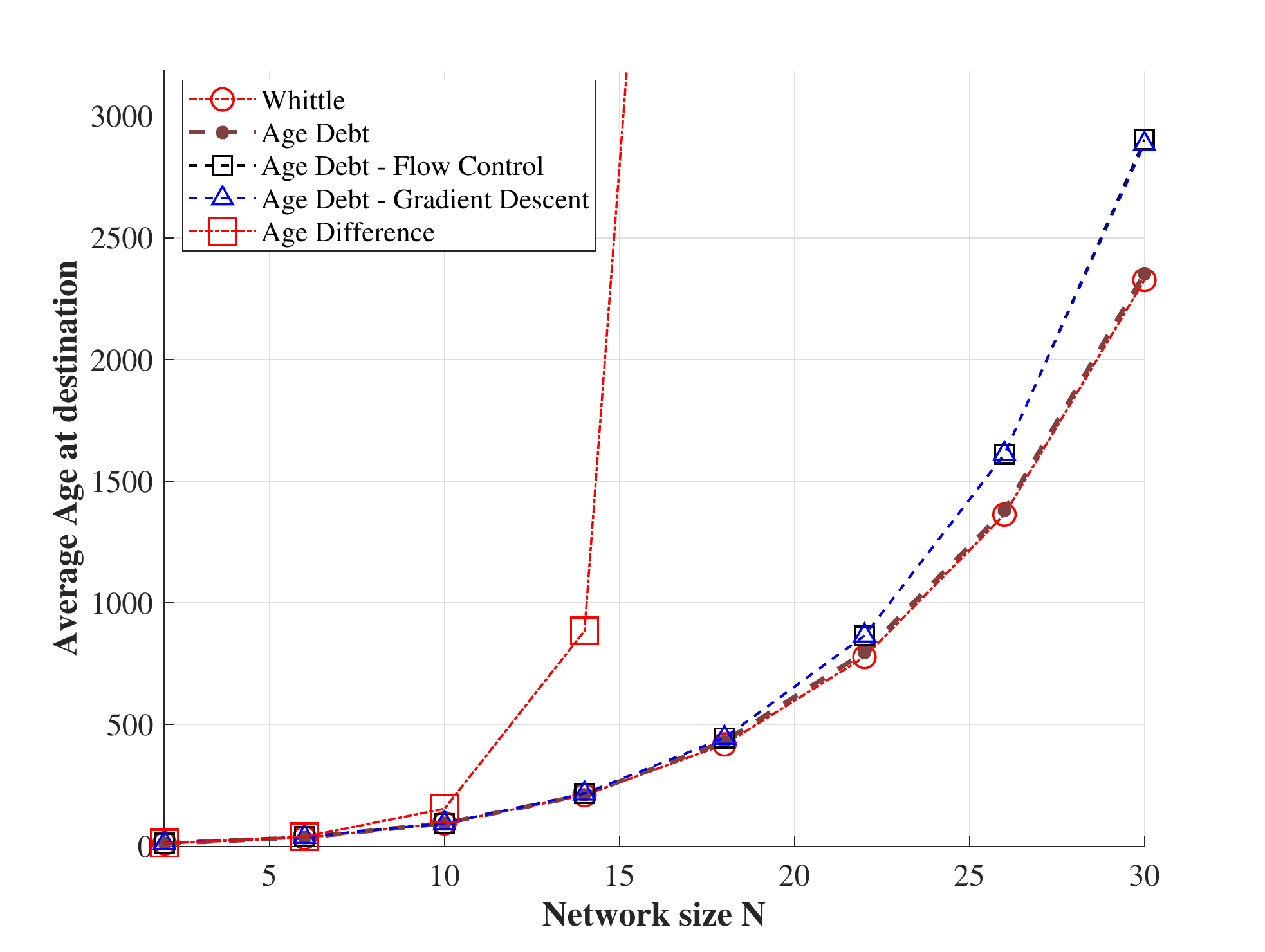}
	\caption{Functions of Age minimization in broadcast networks with reliable channels}
	\label{fig:single_func_rch}
\end{figure}
Next, we consider general functions of age minimization in the single-hop wireless broadcast setting. There are $N$ nodes in the network and the cost of AoI for each node is chosen from the set of functions $\{15A(t), e^{A(t)}, (A(t))^2 \text{ and }  (A(t))^3\}$. Figure \ref{fig:single_func_rch} plots the performance of the age-debt policy and its variants along with the age difference policy and the Whittle index policy proposed in \cite{tripathi_2019_whittle}. As for the linear AoI case, we observe that age debt is able to replicate the Whittle policy's performance when provided its average cost as the target vector. The flow control and gradient descent variants are also only a small gap away in performance without knowing $\bm{\alpha}$ beforehand. On the other hand, the age difference policy performs much worse and the AoI cost rapidly grows very large even for moderate $N$. This is because the age difference policy is not designed to handle general AoI cost functions, so even though it tries to keep the AoIs small for all nodes, their actual impact to cost can become very large. It was also shown in \cite{tripathi_2019_whittle} that even the optimal stationary randomized policy can have unbounded AoI cost for systems as small as $N=2$, given nonlinear AoI cost functions. So, we do not plot its performance in this scenario.

We also look at the functions of age problem with $N=4$ in more detail. The age cost functions for each node are as follows $f_1(A_1(t)) = 15A_1(t)$, $f_2(A_2(t)) = e^{A_2(t)}$, $f_3(A_3(t)) = (A_3(t))^2$ and $f_4(A_4(t)) = (A_4(t))^3$. First, we use dynamic programming to compute the optimal policy $\pi^{*}$ which minimizes average age cost. The time average age costs under this policy are given by $\alpha^{*}_1 = 45.0, \alpha^{*}_2 = 14.52, \alpha^{*}_3 = 17.20,$ and $\alpha^{*}_4 = 11.0$,  while the total sum cost is 87.72. Using these as target values, we set up debt queues and implement the age-debt policy. 

\begin{figure}
	\centering
	\includegraphics[width=0.99\linewidth]{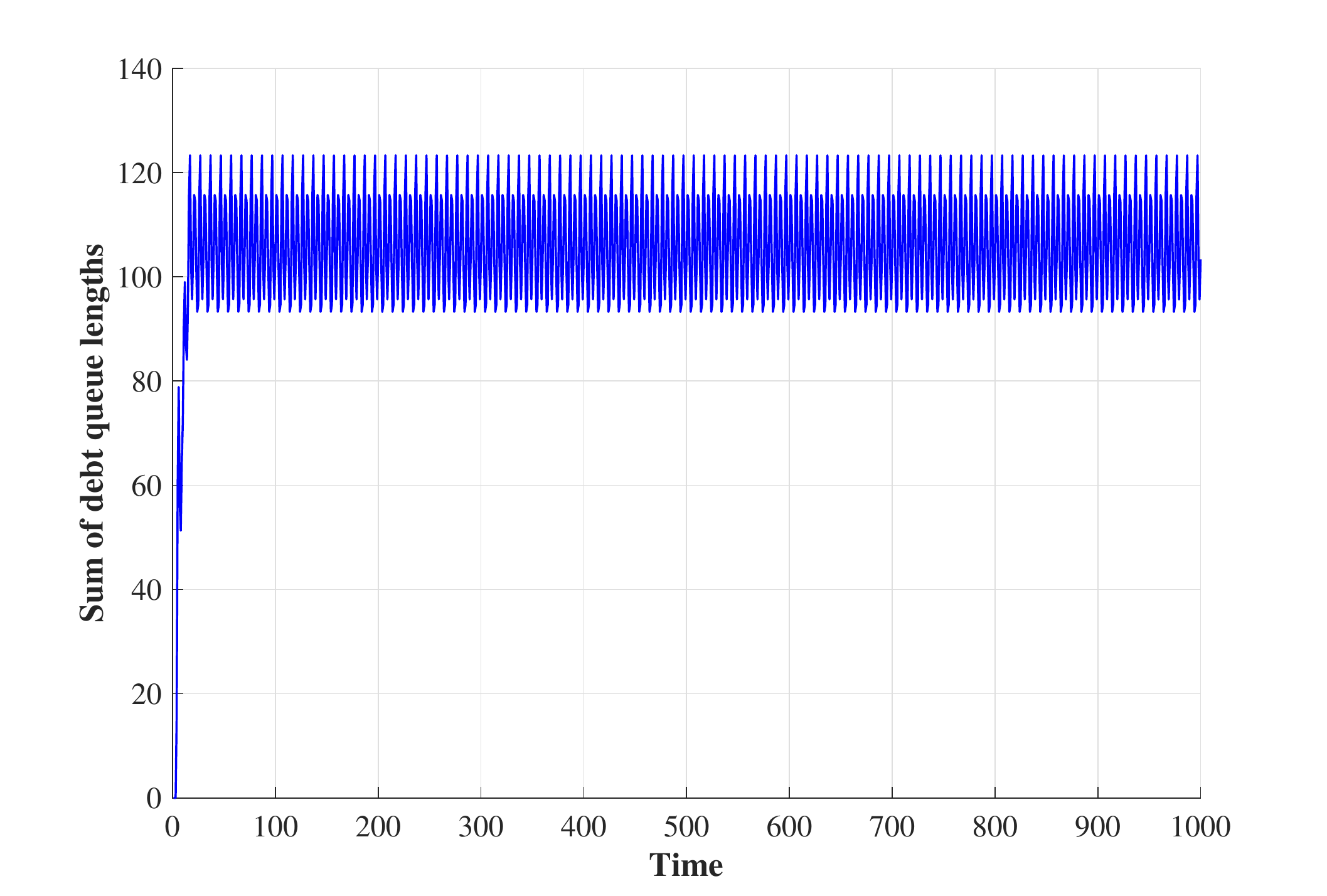}
	\caption{Sum of virtual debt queues vs time}
	\label{fig:dql}
\end{figure}
Figure \ref{fig:dql} plots the sum of the 4 age debt queues $\sum_{i=1}^{4}Q_i(t)$ under the age-debt policy implemented using the optimal $\bm{\alpha}^{*}$ from above. We observe that the age debt policy indeed stabilizes the debt queues since queue lengths don't grow with time. As a corollary, it also achieves age cost optimality in this setting. On the other hand, the Whittle index policy from \cite{tripathi_2019_whittle} achieves a total sum cost of 88.34, a fixed but small distance away from the optimal cost of 87.72. This suggests that age-debt might be a way to achieve exact optimality instead of near optimality when access to $\bm{\alpha}^{*}$ is available.

\begin{figure}
	\centering
	\includegraphics[width=0.99\linewidth]{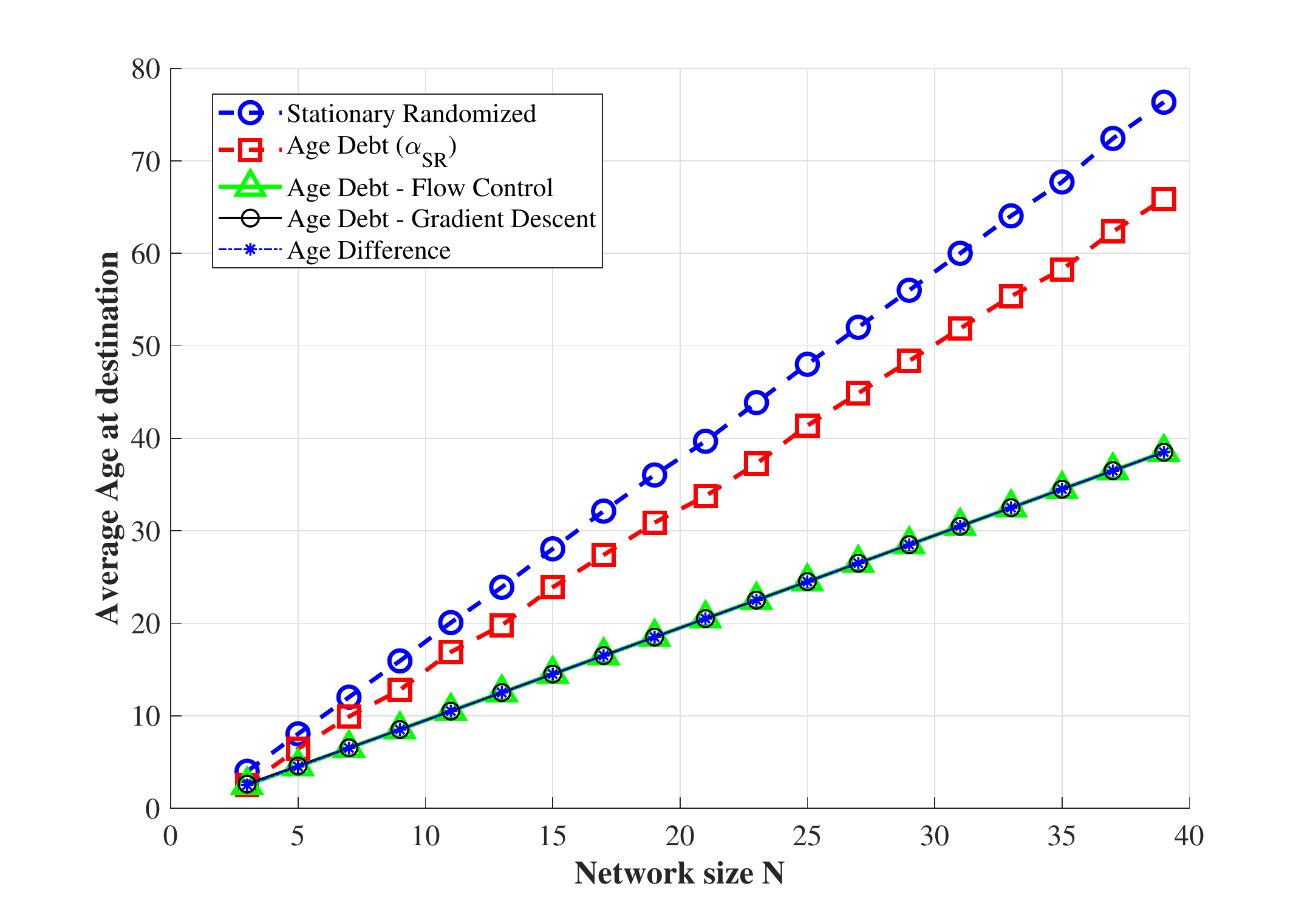}
	\caption{A single unicast flow on a line network (neighboring nodes interfere)}
	\label{fig:unicast_line}
\end{figure}
Next, we consider scheduling for a single unicast flow on the line network. Consider $N$ nodes arranged in a line network from 1 to $N$. Node $1$ wants to sent packets to node $N$, however not all nodes can transmit simultaneously. We consider a simple interference constraint - in any given time-slot either all even numbered nodes or all odd numbered nodes can forward packets. This ensures that no two adjacent nodes send interfering transmissions. Figure \ref{fig:unicast_line} plots the performance of age-debt and its flow-control and gradient-descent variations along with the optimal stationary randomized policy proposed in Section \ref{sec:srp} and the age difference policy proposed in Section \ref{sec:age-diff}. We observe that age-debt outperforms the stationary randomized policy despite using its average cost $\bm{\alpha}_{SR}$ as the target vector. The dynamic variants of age-debt significantly outperform the stationary randomized policy and match the performance of the age-difference policy. We also note that the gap in performance would increase in settings  with multiple flows and paths available which age-debt can utilize for routing, unlike the stationary randomized and age difference approaches.

\begin{figure}
	\centering
	\includegraphics[width=0.99\linewidth]{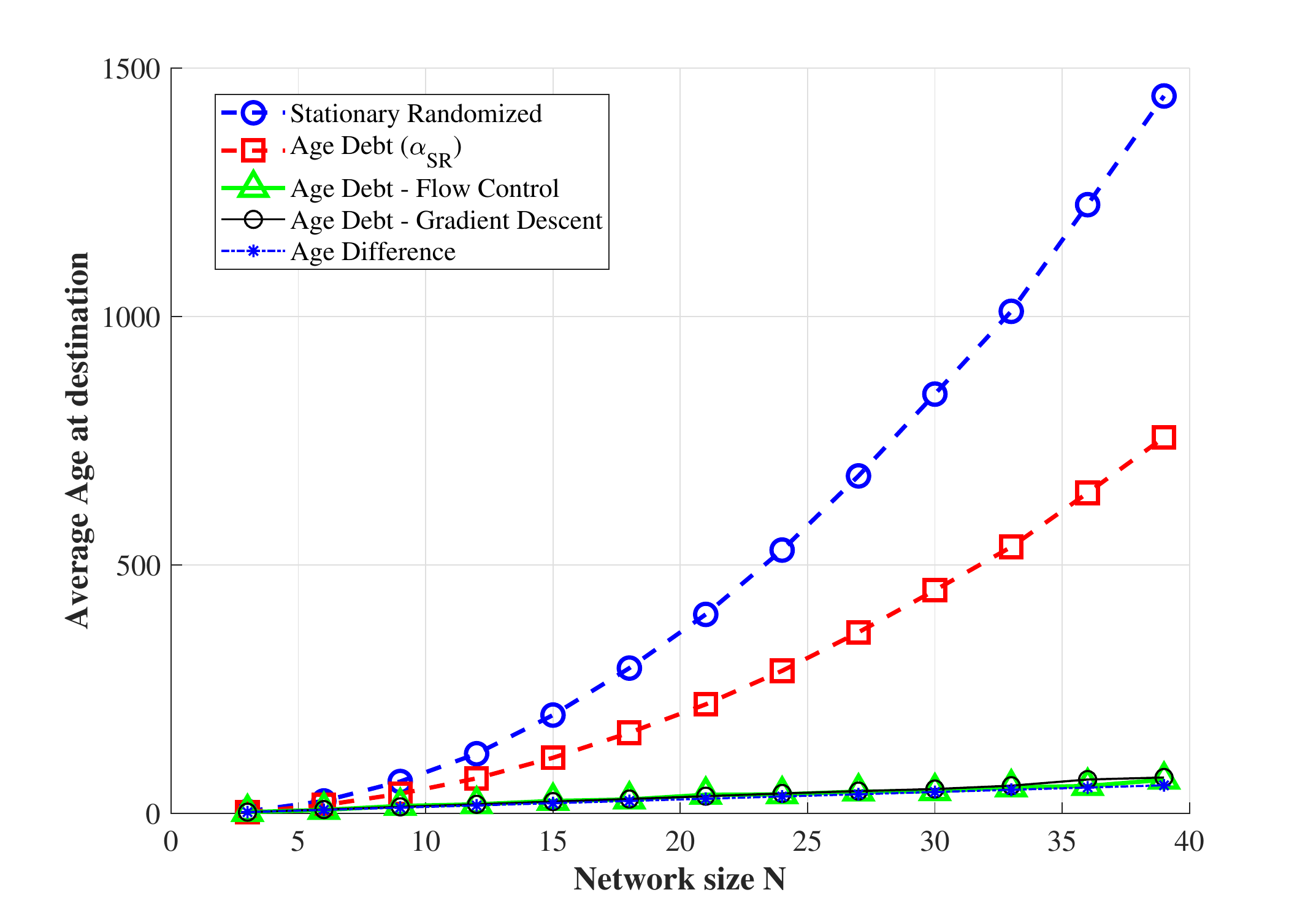}
	\caption{A single unicast flow on a line network (all nodes interfere)}
	\label{fig:unicast_line_s}
\end{figure}
We also consider a different kind of interference constraint in the same line network example. Now, all nodes interfere with one another, and only one node can transmit successfully in any given time-slot. We plot the performance of the optimal stationary randomized policy, the age-debt policy (provided $\bm{\alpha}_{SR}$),  our age-debt variants without any knowledge of $\bm{\alpha}$ and the age difference policy against the number of nodes in the system in Figure \ref{fig:unicast_line_s}. We again observe a large gap in performance between the optimal randomized policy and our proposed methods. This is consistent with the line network AoI analysis from Section \ref{sec:age-diff}, where we showed that the best stationary randomized policy has performance that is $O(N^2)$ while the age difference policy has performance $O(N)$.  We also observe that the age-debt variants match the performance of the age difference policy, which can be shown to be exactly optimal in this single source line network setting.

\begin{figure}
	\centering
	\includegraphics[width=1.05\linewidth]{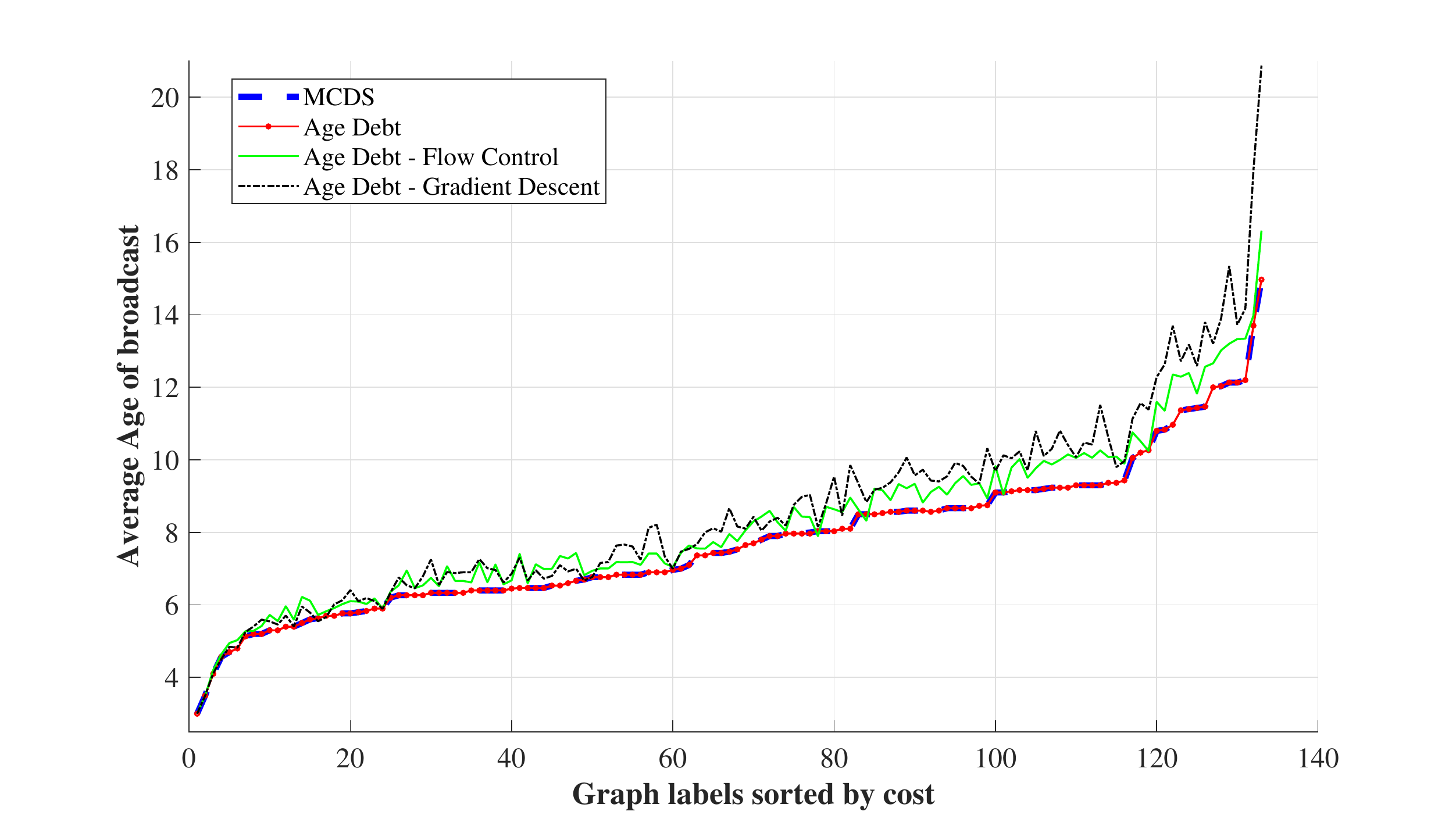}
	\caption{Broadcast flows in multihop networks with 5 and 6 nodes}
	\label{fig:multi_equal}
\end{figure}
Finally, we consider average age minimization for all-to-all broadcast flows in multihop networks similar to \cite{farazi_2018_multihop}. Note that this is a broadcast setting that requires both scheduling and routing decisions to be made, so we cannot use the stationary randomized or age difference policies developed in Sections \ref{sec:srp} and \ref{sec:age-diff}. We consider all possible connected network topologies with 5 or 6 nodes (a total of 133 graphs). Figure \ref{fig:multi_equal} plots the performance of the age-debt policy and its variants along with the near optimal minimum connected dominating set (MCDS) based scheme proposed in \cite{farazi_2018_multihop} for each of these networks. The x-axis represents the graph labels numbered from 1 to 133, sorted according to the average age achieved by the MCDS scheme. 

We observe that age-debt achieves the same performance as the MCDS scheme when provided its average cost as the target vector. Further, age-debt with flow control achieves performance that is very close to that of the MCDS scheme without requiring knowledge of $\bm{\alpha}$. Importantly, the MCDS scheme can only be applied to this setting of all-to-all broadcast with one node transmitting at a time. Further, computing the optimal schedule using the MCDS scheme requires finding minimum size connected dominating sets, the complexity of which grows exponentially in the number of nodes.

\begin{figure}
	\centering
	\includegraphics[width=1.05\linewidth]{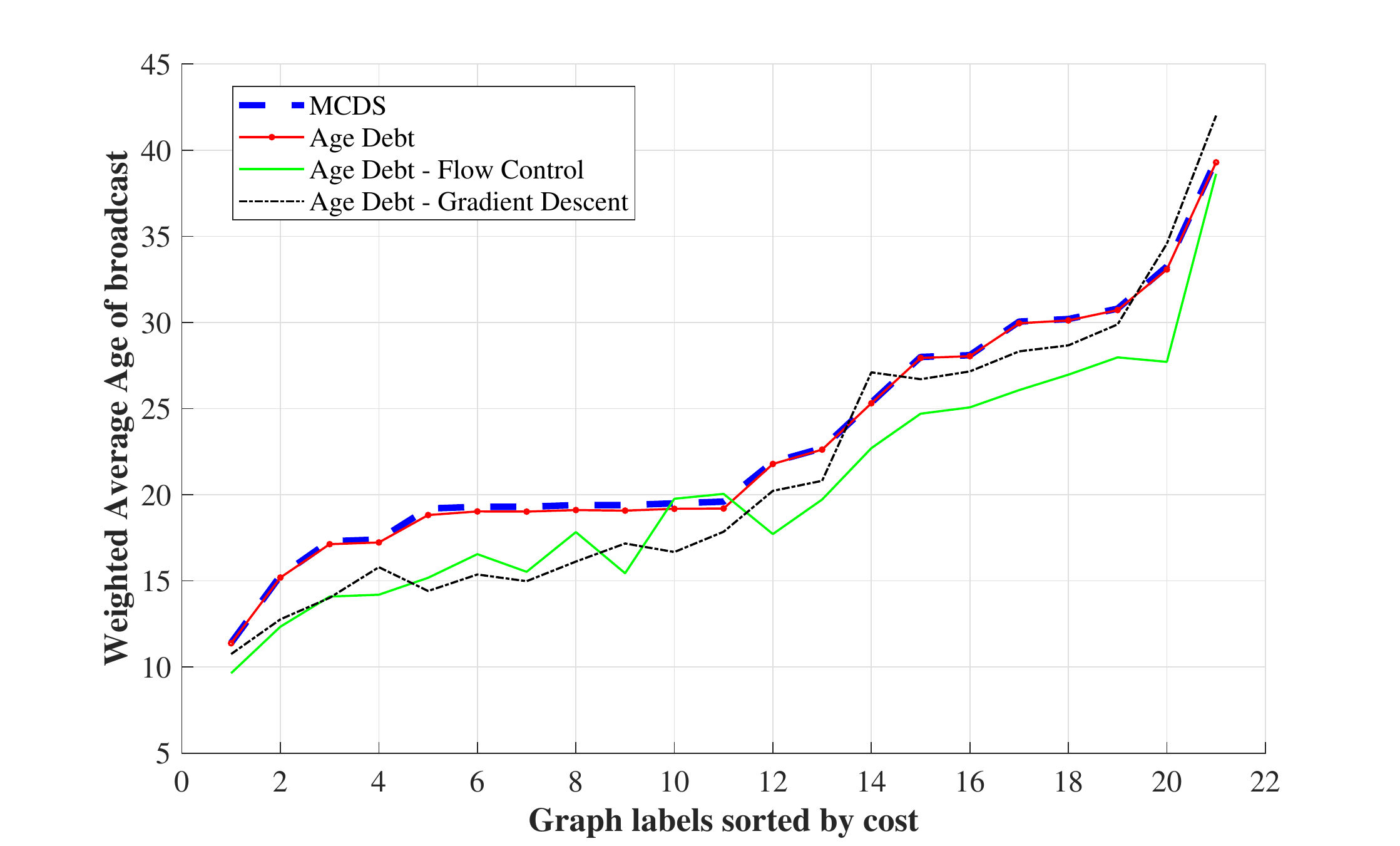}
	\caption{Weighted Age minimization of broadcast flows in multihop networks with 5 nodes}
	\label{fig:multi_w}
\end{figure}
We also consider the same broadcast setting but now with weighted-sum AoI as the minimization objective instead of just AoI. We consider all possible connected graphs with 5 nodes (21 in total). We set the importance weight of one node to $15$ (giving it a higher priority) and the rest of the 4 nodes to $1$. Figure \ref{fig:multi_w} plots the performance of the MCDS scheme along with age-debt and its variants. As expected, age-debt policy replicates the performance of the MCDS scheme since it is provided the average age-cost realized by MCDS as the target. Interestingly, flow-control outperforms MCDS since it is able to adapt to a better target $\bm{\alpha}$ in the presence of weights and asymmetry. This is consistent with the fact that the MCDS scheme is not designed for minimizing weighted-sum AoI. It also highlights the relative ease with which age-debt can be adapted to weights and general AoI cost functions.

Note that the complexity of implementing the flow-control scheme is polynomial in the network size per time-slot. This suggests that age-debt and its variants are a good candidate for low complexity near optimal age scheduling in general networks.

We also observe that the flow control variant of age-debt is the method of choice in the absence of known $\bm{\alpha}$. During our experiments, we found that the gradient descent variant has parameters that are hard to configure for networks of different sizes and takes a long time to converge. The flow-control method has just two parameters $V$ and $\alpha_{\text{max}}$ that are relatively easy to configure and do not require any time for convergence.

\section{Conclusions}
\label{sec:conclusion}
We considered the problem of minimizing age-of-information freshness metrics for general multi-hop networks and proposed three classes of policies - stationary randomized, age-difference and age-debt. Through analysis and simulations, we compared the performance of these three policies with the best known policies previously known in literature for a wide variety of settings. Directions of future exploration involve 1) proving performance bounds for age-debt and its variants, and 2) considering distributed implementation, stochastic arrivals and time-varying network topologies.

\section{Acknowledgments}
This work was supported by NSF Grants AST-1547331, CNS-1713725, and
CNS-1701964, and by Army Research Office (ARO) grant number W911NF17-1-0508. We also thank Shahab Farazi for sharing with us an implementation of the MCDS scheme from \cite{farazi_2018_multihop}. 
%\color{blue} write conclusion. next steps - stochastic arrivals, distributed implementation, time-varying topologies, and performance bounds. Thank Farazi for sharing MCDS broadcast code and implementation.\color{black}

%We derived AoI optimal stationary scheduling policies, in which links are activated according to a stationary probability distribution.

%We first considered a simple line network, with a single source-destination pair, and showed that the age is a simple separable convex function of link activation frequencies. We then used this result to prove an important \emph{separation principle} for the general multi-hop network with $\mathcal{R}$ source-destination pairs. The separation principle states that the optimal stationary policy for the multi-hop network can be obtained by solving an equivalent problem in which all source-destination pairs are just a single-hop away. 

\bibliographystyle{ieeetr}
\bibliography{aoi-bib,books-bib}

\appendix

\subsection{Proof of Theorem~\ref{thm:age_lineNet}}
\label{pf:thm:age_lineNet}
For any stationary randomized policy, we show that
\begin{equation}
\label{eq:no2}
\EX{\Age{N}{}(t)} = \sum_{e \in E}\frac{1}{\CSprob{e}{}\linkActFreq{e}{}{}} = \sum_{j=2}^{N}\frac{1}{\CSprob{\edgel{j}}{}\linkActFreq{\edgel{j}}{}{}},
\end{equation}
where \linkActFreq{\edgel{j}}{}{} is the link activation frequency for link \edgel{j} and \CSprob{\edgel{j}}{} is the probability that the state of the link \edgel{j} is 1. The result (in~\eqref{eq:no1}) follows from~\eqref{eq:no2}.

Let $\pi$ be a stationary randomized policy, and \linkActFreq{\edgel{j}}{}{} be the link activation frequency for link \edgel{j} under policy $\pi$, for all $j \in \{2, \ldots L\}$. Let $\alpha_j \triangleq \CSprob{\edgel{j}}{}\linkActFreq{\edgel{j}}{}{}$.
Further, let $\tau_{j}(t)$ denote the last instance, when a successful transmission occurred over link \edgel{j}. For example, if link activations occurred at time slots $2, 10, 14,$ and $21$ then $\tau_{j}(t) = 10$ for all $t = 11, 12, 13,$ and $14$.

Since $\pi$ is a stationary randomized policy, the (successful) inter-transmission times must be geometrically distributed with mean $\frac{1}{\alpha_j}$, for link \edgel{j}. The memoryless property, therefore, implies that
\begin{equation}\label{eq:ndtv}
\pr{\tau_{j}(t) = t - s} = \alpha_{j}\left( 1 - \alpha_{j}\right)^{s-1},
\end{equation}
for all $s = 1, 2, \ldots$. Thus, $\tau_{j}(t)$ has the same distribution as $t - X_{\edgel{j}}$, where $X_{\edgel{j}}$ is a geometrically distributed random variable given by
\begin{equation} \label{eq:loc_z0}
\pr{X_{\edgel{j}} = s} = \alpha_{j}\left( 1 - \alpha_{j} \right)^{s-1},
\end{equation}
for all $s \in \{1, 2, \ldots \}$, with mean
\begin{equation}\label{eq:loc_z1}
\EX{X_{\edgel{j}}} = \frac{1}{\alpha_j},
\end{equation}
for all the links \edgel{j} in the network.

Node $1$ being the active source, its age is always $0$. Consider age at node $2$. Since the source node always transmits fresh information, the age of node $2$ is the just the time elapsed since the last successful transmission over the link \edgel{2}. This is given by
\begin{equation}
\label{eq:a1}
\Age{2}{}(t) = t - \tau_{2}(t),
\end{equation}
as $\tau_{2}(t)$ was the last time a fresh update packet was sent to node $1$.

Next, consider age at node $3$.
By Lemma~\ref{lem:age_evol}, whenever a successful transmission occurs over link \edgel{3}, node $3$ resets its age to node $2$'s age $\Age{2}{}(\cdot)$. Thus, $\Age{3}{}(t)$ is given by
\begin{equation}
\label{eq:a2}
\Age{3}{}(t) = t - \tau_{3}(t) + \Age{2}{}\left( \tau_{3}(t)\right),
\end{equation}
where $\Age{2}{}\left( \tau_{3}(t)\right)$ is the age of node $2$ at the time of the last successful transmission over link \edgel{3}, namely $\tau_{3}(t)$, and $t - \tau_{3}(t)$ is the time elapsed since then.
Substituting~\eqref{eq:a1} in~\eqref{eq:a2}, we obtain
\begin{align}
\Age{3}{}(t) &= t - \tau_{3}(t) + \left[ \tau_{3}(t) - \tau_{2}\left( \tau_{3}(t)\right) \right], \\
&= t -  \tau_{2}\left( \tau_{3}(t)\right).
\end{align}
Iterating this over $j$ links, we get
\begin{equation}\label{eq:loc_z10}
\Age{j}{}(t) = t - \tau_2\left( \tau_3\left( \cdots \tau_j\left( t \right) \cdots  \right) \right),
\end{equation}
for all $t$ and nodes $j \in \{2, 3, \ldots N\}$.
Taking expectation, we get
\begin{align}
\EX{\Age{j}{}(t)} &= t - \EX{\tau_2\left( \tau_3\left( \cdots \tau_j\left( t \right) \cdots  \right) \right)}, \\
&= \sum_{i=2}^{j} \frac{1}{\alpha_i},
\end{align}
where the last equality follows from the following Lemma~\ref{lem:tau}, and  substituting $j = N$ we obtain~\eqref{eq:no2}, and also, the result.
\begin{lemma}\label{lem:tau}
$\tau_{j}(t)$ for $j \in \{2, 3, \ldots N\}$ satisfy
\begin{equation}\label{eq:node1}
\EX{\tau_2\left( \tau_3\left( \cdots \tau_j\left( t \right) \cdots  \right) \right)} = t - \sum_{i=2}^{j}\frac{1}{\alpha_i}.
\end{equation}
\end{lemma}
\begin{IEEEproof}
%Proof of Lemma~\ref{lem:tau} is given in Appendix~\ref{pf:lem:tau}
%\subsection{Proof of Lemma~\ref{lem:tau}}
%\label{pf:lem:tau}
From~\eqref{eq:ndtv}, we know that
\begin{equation}
\label{eq:zee}
\tau_{i}(t) \overset{d}{=} t - X_{\edgel{i}},
\end{equation}
where $X_{\edgel{i}}$ is a geometrically distributed random variable given by~\eqref{eq:loc_z0},
and therefore,
\begin{equation}
\label{eq:zee_ex1}
\EX{\tau_{i}(t)} = \EX{t - X_{\edgel{i}}} = t - \frac{1}{\alpha_i},
\end{equation}
for all $i \in \{2, 3, \ldots N\}$.
Now from~\eqref{eq:zee}, we can obtain
\begin{equation}
\label{eq:zee2}
\tau_{i}\left(\tau_{i+1}(t)\right) \overset{d}{=} \tau_{i+1}(t) - X_{\edgel{i}},
\end{equation}
and therefore,
\begin{align}\label{eq:zee_ex2}
\EX{\tau_{i}\left(\tau_{i+1}(t)\right)} &= \EX{\tau_{i+1}(t)} - \EX{X_{\edgel{i}}}, \\
&= t - \frac{1}{\alpha_{i+1}} - \frac{1}{\alpha_i},
\end{align}
where the last equality follows from~\eqref{eq:zee_ex1} and~\eqref{eq:loc_z1}. Iterating this $u$ times we obtain:
\begin{equation}\label{eq:aSDouf1}
\EX{\tau_{i}\left(\tau_{i+1}( \cdots \tau_{i+u}(t)\cdots)\right)} = t - \sum_{l=i}^{i+u}\frac{1}{\alpha_l}.
\end{equation}
Substituting $i = 2$ and $i+u = j$ we get the result.
\end{IEEEproof}

\subsection{Proof of Lemma~\ref{lem:equi}}
\label{pf:thm:equi}

The optimization problem~\eqref{eq:age_opt} is given by
\begin{align}
%\label{eq:age_opt}
\begin{aligned}
& \underset{\mathbf{x} \in \setX}{\text{Minimize}}
& & \sum_{k \in [K]} \sum_{e \in \Path{k}} \frac{w^{k}}{\CSprob{e}{}\linkActFreq{e}{}{k} }, \\
%& \text{subject to} & &  \sum_{k = 1}^{K} \linkActFreq{\edgel{j}}{}{k} \leq \linkActFreqSum{\edgel{j}}{}~\forall e \in E \\
&\text{subject to} && \mathbf{f} = M \mathbf{x}.%\linkActFreqVec{k} = Q^{k}\mathbf{f}~\forall~k = 1, 2, \ldots K,~\text{and}\\
%&&& \mathbf{f} = M \mathbf{x}.
\end{aligned}
\end{align}
Consider the set of tuples $ H = \{ (k,e), \forall k \in [K], \forall e \in p^{k} \}$. The objective in~\eqref{eq:age_opt} involves one term for each element in $H$. To create a new single-hop network with the same age minimization problem, we create a source-destination pair corresponding to each element $h = (k,e) \in H$. The link reliability of the edge for the source-destination pair $h$ is given by $\gamma_{e}$, while the weight for AoI at the destination is given by $w^{k}$. The set of feasible activations for the original multihop network $\mathcal{A}$ contains interference free choices of flow and edge activations of the form $\{(k_1, e_1), (k_2, e_2), ...\}$. For our new multihop network, we simply translate this to general interference constraints. For example, if $\{(k_1, e_1), (k_2, e_2)\} \in \mathcal{A}$, then the sources corresponding to $(k_1, e_1)$ and $(k_2, e_2)$ can attempt to transmit simultaneously in the new single-hop network without interference. Now, we can construct the single-hop age minimization problem over this new network using \eqref{eq:shap} as below:    
\begin{align}
%\label{eq:shap}
\begin{aligned}
& \underset{\mathbf{x} \in \setX}{\text{Minimize}}
& & \sum_{(k,e) \in H} \frac{w^{k}}{\CSprob{e}{}\linkActFreq{e}{}{k} }, \\
%& \text{subject to} & &  \sum_{k = 1}^{K} \linkActFreq{\edgel{j}}{}{k} \leq \linkActFreqSum{\edgel{j}}{}~\forall e \in E \\
&\text{subject to} && \mathbf{f} = M \mathbf{x}.
\end{aligned}
\end{align}
Note that this is identical to \eqref{eq:age_opt}, which completes the proof.

\subsection{Proof of Lemma~\ref{lem:age-diff}}
\label{pf:lem:age-diff}
The per time-slot cost in \Aave (in~\eqref{eq:aveAoI_plusCk}) is
\begin{equation}\label{eq:Ct}
C(t) = \sum_{k=1}^{K}\sum_{j \in C_k\cup D_k} \weight{j}{k} \Age{j}{k}(t).
\end{equation}
Note that the age evolution in~\eqref{eq:AoI_evolution_mh} can be re-written as
\begin{multline}\label{eq:age_evolve_new}
\Age{j}{k}(t+1) = 1 + \Age{j}{k}(t) \\
- \sum_{i} \Tx{i}{j}{k}(t)\CS{i}{j}(t) \pos{ \Age{j}{k}(t) - \Age{i}{k}(t)},
\end{multline}
for all $j \in C_k \cup D_k$, $i \in \{k\}\cup C_k$ that are neighbors of $j$; in~\eqref{eq:age_evolve_new} $\pos{x}$ is used to denotes $\max\{x, 0\}$.

From~\eqref{eq:Ct}-\eqref{eq:age_evolve_new}, we deduce that the per time-slot cost difference will be
\begin{multline}
\label{eq:luiwe1}
C(t+1) - C(t) = \sum_{k = 1}^{K}\sum_{j \in C_k \cup D_k} \weight{j}{k} \\
- \sum_{\edge{i}{j} \in \bar{E}}\sum_{k=1}^{K} \Tx{i}{j}{k}(t)\CS{i}{j}(t) \pos{ \Age{j}{k}(t) - \Age{i}{k}(t)},
\end{multline}
where $\bar{E}$ that are present in the subgraph induced by all the source, commissioned, and destination nodes. $\bar{E}$ accounts for all the links on which updates will be forwarded for at least one flow.
Writing~\eqref{eq:luiwe1} in terms of weights $\ADweights{i}{j}{k}(t)$, we have:
\begin{equation}
\label{eq:luiwe2}
C(t+1) - C(t) = \sum_{k = 1}^{K}\sum_{j \in C_k \cup D_k} \weight{j}{k}
- \sum_{\edge{i}{j} \in \bar{E}}\sum_{k=1}^{K} \ADweights{i}{j}{k}(t).
\end{equation}
The age-difference policy maximizes the sum
\begin{equation}
\sum_{\edge{i}{j} \in \bar{E}}\sum_{k=1}^{K} \ADweights{i}{j}{k}(t).
\end{equation}
As a result, it minimizes $C(t+1)$, given all occurrences till time $t$. This shows that the age-difference policy is a myopic policy for the average age defined in~\eqref{eq:aveAoI_plusCk}.

\subsection{Proof of Lemma \ref{lem:debt_equi}}
\label{pf:lem_equi}
We will prove this under the assumption that the AoI cost functions $g^{k}_{j}(\cdot)$ are upper-bounded by a fixed constant $D$ for every source-destination pair $(k,j)$. This is a mild assumption because $D$ can be set to a very high value (in the order of years) which will never be attained in practical systems under any reasonable policy.

We note that the arrival process to the debt queue $Q^{k}_{j}(t)$ is given by the effective age process $B^{k}_{j}(t)$, while the departures in every time-slot are just $\alpha^{k}_{j}$. Using the boundedness assumption, both arrivals and departures are strictly upper-bounded by $D$. The result immediately follows from Theorem 2(c) in \cite{neely_2010_stability_definitions} which relates mean-rate stability of a queue to time-averages of the arrival and departure processes.

\subsection{Proof of Lemma \ref{lem:drift_exp}}
\label{pf:drift}
The debt queues in this setting evolve as follows:
\begin{equation}
Q_i(t+1) = \bigg[Q_i(t) + g_i(A_i(t+1)) - \alpha_i\bigg]^{+}, \forall i.
\end{equation}
The AoI evolves as:
\begin{equation}
A_i(t+1)
\begin{cases}
A_i(t)+1, & \text{if } U_{e_i}(t)S_{e_i}(t) = 0 \\
1, &  \text{if } U_{e_i}(t)S_{e_i}(t) = 1.
\end{cases}
\end{equation}
Here $S_{e_i}(t) = 1$ i.i.d. with probability $\gamma_i$ in every time-slot.

Let $\Delta(t) \triangleq L(t+1) - L(t)$. Then,
\begin{equation}
\begin{split}
\mathbb{E}[\Delta(t)] & = \sum_{i} \mathbb{E}\bigg[ (Q_i(t+1))^2 - (Q_i(t))^2 \bigg]\\
& \leq \sum_i \mathbb{E}\bigg[ \alpha_i^2 - 2\alpha_i Q_i(t) + (g_i(A_i(t+1)))^2 + \\ & 2Q_i(t)g_i(A_i(t+1)) - 2\alpha_i g_i(A_i(t+1)) \bigg]\\
& \leq \sum_i \bigg[ D^2 + 2 Q_i(t) (\mathbb{E}[g_i(A_i(t+1))] - \alpha_i) \bigg]
\end{split}
\end{equation}

The first inequality follows from the evolution of debt queues. The second inequality follows from the boundedness assumption on $g_{i}(\cdot)$, i.e. $g_i(h) \leq D, \forall h$. Now, we will minimize the RHS of the expression above. We can drop the term $D^2$ since it is a constant.
\begin{equation}
\begin{split}
&\underset{\pi(t) \in 1,...,N-1}{\operatorname{argmin}} \sum_i  Q_i(t) \big(\mathbb{E}[g_i(A_i(t+1))] - \alpha_i\big) \\
= & \underset{\pi(t) \in 1,...,N-1}{\operatorname{argmin}} \sum_i  Q_i(t) \mathbb{E}[g_i(A_i(t+1))] \\
= & \underset{j \in 1,...,N-1}{\operatorname{argmin}} \bigg[\sum_i \bigg( Q_i(t) g_i(A_i(t)+1) \bigg) + \\
& ~~~~~~\gamma_j Q_j(t) (g_j(1) - g_j(A_j(t) + 1))\bigg]\\
= & \underset{j \in 1,...,N-1}{\operatorname{argmax}} \bigg[ \gamma_j Q_j(t) \big(g_j(A_j(t) + 1) - g_j(1)\big) \bigg]
\end{split}
\end{equation}
The first equality follows since $Q_i(t)\alpha_i$ does not depend on the scheduling decision $\pi(t)$. The second equality follows from the evolution of AoI given $\pi(t)=j$. The third equality follows since the summation term does not depend on the scheduling choice $j$. This completes the proof.

\end{document}